\documentclass{amsart}

\usepackage{amsmath,amsfonts,amssymb,amsthm,graphicx,cite}
\usepackage{tikz-cd}

\theoremstyle{remark}

\usepackage{xcolor}

\usepackage{hyperref}

\newcommand{\om}{\omega}
\newcommand{\C}{\mathbb{C}}
\newcommand{\R}{\mathbb{R}}
\newcommand{\Z}{\mathbb{Z}}
\newcommand{\ee}{\mathrm{e}}
\newcommand{\ii}{\mathrm{i}}
\newcommand{\cC}{\mathcal{C}}
\newcommand{\cE}{\mathcal{E}} 

\begin{document}

\title{Elliptic Integrable Systems and Special Functions}

\author{Martin Halln\"as}
\email{hallnas@chalmers.se}
\address{Department of Mathematical Sciences, Chalmers University of Technology and the University of Gothenburg, SE-412 96 Gothenburg, Sweden}

\author{Edwin Langmann}
\email{langmann@kth.se}
\address{Department of Physics, KTH Royal Institute of Technology, SE-106 91 Stockholm, Sweden} 

\date{\today}

\begin{abstract}
We discuss elliptic quantum Calogero-Moser-Sutherland models, including their relativistic generalizations due to Ruijsenaars and van Diejen, and the relations of these models to classes of special functions developed and explored in recent and on-going research efforts. 
We give an introduction to these models and corresponding special functions aimed at non-experts. 
We also describe a few open problems in the field. 
\end{abstract}


\maketitle

\section*{Prologue}
There are many integrable systems related to elliptic functions, including models in statistical physics (e.g.\ the eight-vertex model), soliton theory (e.g.\ the periodic KdV equation), mechanics (e.g.\ the Kovalevskaya top), and quantum field theory (e.g.\ certain supersymmetric Yang-Mills gauge theories). We cannot cover all these systems,  and we therefore interpret {\em elliptic integrable system} in a narrow sense as {\em elliptic quantum Calogero-Moser-Sutherland models}, including their relativistic generalizations due to {\em Ruijsenaars} and {\em van Diejen}; see e.g.\ the chapter on Calogero-Moser-Sutherland models in this encyclopedia by Halln\"as (2024) \cite{hallnas2024} for an introduction to these quantum systems referred to as {\bf Calogero-Moser-Ruijsenaars (CMR) models} in the following. 

In order not to overload the main text with footnotes, we collect  complementary Remarks~\ref{app:C}.\ref{rem:1}  etc.\ in Appendix~\ref{app:C}; these can be skipped without loss of continuity.
 
\section{Introduction}\label{sec:intro}  
Hermite polynomials providing exact eigenfunctions of the quantum harmonic oscillator are probably the most famous example of the close relation between special functions and exactly solvable systems in quantum physics. 
More recent such examples include the Jack polynomials  providing the exact eigenfunctions of the trigonometric quantum Calogero-Moser system first solved by Sutherland, and the Macdonald polynomials related in a similar way to its relativistic generalization known as the trigonometric Ruijsenaars model (Jack (1970), Sutherland (1972), Macdonald (1988), and Ruijsenaars (1987) \cite{ruijsenaars1987}; see Remark~\ref{app:C}.\ref{rem:1}).

Elliptic CMR systems are quantum integrable systems with eigenfunctions that are more complicated than the ones appearing in the trigonometric case, and these functions are subjects of current research. 
One ambition in this research is to turn elliptic CMS models from {\em integrable} to {\em exactly solved}, and this requires to develop a better understanding and explicit constructions of new classes of special functions generalizing the  polynomials  appearing in the trigonometric case.
This chapter gives an introduction to these developments. 
More specifically, our aim is four-fold: (i) introduce the family of elliptic CMR-type models to non-experts, (ii)~give an introduction to some known results about eigenfunctions of elliptic CMR-type systems, (iii) guide the reader to the literature on eigenfunctions of elliptic CMR-type systems, (iv) describe a few outstanding open question in the field. 

CMR models come in four variants: (I) rational,  (II) hyperbolic, (III) trigonometric, (IV) elliptic, depending on the kind of function used in the definition of the model; elliptic CMR models are the most general case (IV), i.e., (I), (II) and (III) are limiting cases of (IV). 
In particular, to define elliptic CMR models in Section~\ref{sec:CMR}, we use the following variant of the Weierstrass elliptic $\wp$-function with half-periods $(\ell,\ii\delta)$, 
\begin{align}\label{wp1}  
\wp_1(x) = \sum_{n\in\Z} \frac{(\pi/2\ell)^2}{\sin^2(\pi(x-2 n\ii \delta)/2\ell)}
\end{align} 
where $\ell>0$ and $\delta>0$ ($\ii=\sqrt{-1}$ and $x\in\C$), together with the theta function
\begin{align}\label{theta1}  
\vartheta_1(x) & = 2\sin(\pi x/2\ell)\prod_{n=1}^\infty(1-\ee^{\pi\ii (x+2n\ii \delta)/\ell})(1-\ee^{-\ii\pi (x-2n\ii \delta)/\ell})
\end{align} such that $\wp_1(x)=-\partial_x^2\ln \vartheta_1(x)$; Appendix~\ref{app:ell} contains details on how these definitions are related to the ones in the classic reference by Whittaker \& Watson (1920) \cite{whittaker1920},  together with a few well-known identities we need. We restrict to $\ell,\delta\in\R_{>0}$ to allow for a physics interpretation of the operators defining elliptic CMR models as quantum mechanical Hamiltonians; many results we describe can be analytically continued to $\ell,\delta\in\C$ such that the half-period ratio $\ii \delta/\ell$ has a positive imaginary part, and these extensions can be interesting from a mathematical point of view.
 
The elliptic function $\wp_1(x)$ is used to define interaction potentials in the non-relativistic elliptic CMR models, while the ratio $\vartheta_1(x)/\vartheta_1(y)$ appears in their relativistic generalizations.  
The pertinent functions for the cases (I), (II) and (III) are obtained as limiting cases $(\ell,\delta)\to(\infty,\infty)$, $\ell\to\infty$ and $\delta\to\infty$ from the ones for case (IV) as follows: 
\begin{align*} 
\wp_1(x) \to \begin{cases} 1/x^2 & \text{(I)} \\
(\pi/2\delta)^2/\sinh^2(\pi x/2\delta) & \text{(II)}\\
(\pi/2\ell)^2/\sin^2(\pi x/2\ell) & \text{(III)}
\end{cases}
\end{align*} 
and 
\begin{align*}
\vartheta_1(x)/\vartheta_1(y) \to \begin{cases} x/y & \text{(I)} \\
\sinh(\pi x/2\delta)/\sinh(\pi y/2\delta) & \text{(II)}\\
\sin(\pi x/2\ell)/\sin(\pi y/2\ell). & \text{(III)}
\end{cases}
\end{align*} 
For later reference, we note that $\vartheta_1(x) = \ii z^{-1/2}\theta(z;p)$ with 
\begin{align}\label{theta} 
\theta(z;p) = (1-z)\prod_{n=1}^\infty(1-p^n z)(1-p^n/z)
\end{align} 
and $z=\exp(\ii\pi x/\ell)$, where 
\begin{align}\label{p} 
p=\ee^{-2\pi\delta/\ell}
\end{align} 
is the elliptic deformation parameter with $p=0$ being the trigonometric case. The theta function in \eqref{theta} is often used in recent mathematical work on elliptic CMR models.

Typically, results in the trigonometric case (III) are well-established, and generalizations to the general elliptic case (IV) are more recent or remain to be found.
Reasons for why such generalizations are interesting from a physics point of view include: (a) while the short-distance behavior of interactions is the same in cases (III) and (IV), the long-distance behavior is richer in the elliptic than in the trigonometric case, (b) the elliptic case allows for non-stationary generalizations which are easier to solve than the standard elliptic cases, (c) the elliptic case includes two different potentials which are qualitatively different: $\wp_1(x)$ and $\wp_1(x+\ii\delta)$ (and similarly in the relativistic case), and this allows for interesting generalizations of elliptic CMR models which do not exist in the trigonometric case,

We explain (b) and (c) in Sections~\ref{sec:nonstationary} and \ref{sec:generalized}, respectively. By (a) we mean that the elliptic models allow to interpolate between long-range interactions for $\delta\to\infty$ (trigonometric case) and short-range interactions for $\delta\to 0$. 
To see this, we first note that cases (III) and (IV) describe interacting particle systems on a circle parametrized by a coordinate in the interval $[-\ell,\ell]$, while cases (I) and (II) describe interacting particle systems on the real line $\R$. 
Moreover, since 
\begin{align*} 
\sum_{n\in\Z} \frac1{(x-2n\ell)^2} = \frac{(\pi/2\ell)^2}{\sin^2(\pi x/2\ell)}, 
\end{align*} 
the trigonometric case (III) corresponds to the $2\ell$-periodic version of the rational case (I): one replaces the real line $\R$ by the circle of circumference $2\ell$ using the mirror image method. 
Similarly, the elliptic case (IV) corresponds to the periodic version of the hyperbolic case (II) (see Remarks~\ref{app:C}.\ref{rem:wp1} and \ref{app:C}.\ref{rem:RandS1} for further explanations). 
Finally, cases (I) and (II) are different from each other in that hyperbolic non-relativistic CMR systems describe particles interacting with two-body interaction potentials given by the function $(\pi/2\delta)^2/\sinh^2(\pi x/2\delta)$ which at distances $|x|\gg \delta/\pi$ decays exponentially like $(\pi/\delta)^2\exp(-\pi |x|/\delta)$, while the two body interaction potential in the rational case is $1/x^2$ for all real $x$; thus, the long-distance behavior of the interactions is very different in the hyperbolic and rational cases, and the same is true for the elliptic and trigonometric cases. 

From the point of view of quantum integrability (see e.g. Halln\"as (2024) \cite{hallnas2024}), the hyperbolic (II) and trigonometric (III) cases are equivalent in the sense that one can be obtained from the other by replacing $\ii\delta$ by $\ell$. 
However, when it comes to eigenfunctions of interest in quantum mechanics, cases (II) and (III) are different: while the eigenfunctions of trigonometric CMR models are typically given by polynomials  (with Jack and Macdonald polynomials as famous examples), the corresponding eigenfunctions in the hyperbolic case are usually transcendental (and were obtained in works by Heckman \& Opdam, for example; see Remark~\ref{app:C}.\ref{rem:IIandIII}). 

There is a large zoo of special functions defined by elliptic CMR models; while we emphasize the eigenfunctions of CMR models defined by their physics interpretation as quantum mechanical models, we also mention other kinds of eigenfunctions which are of interest from a mathematical point of view.

Our plan is as follows. 
Section~\ref{sec:CMR} gives an overview over the elliptic CMR systems. 
This is followed by three sections giving an introductions to three different kinds of results on functions defined by elliptic CMR models:
Bethe ansatz solutions (Section~\ref{sec:Bethe}), solutions by integrals  (Section~\ref{sec:integrals}), and perturbative solutions (Section~\ref{sec:perturbative}). 
We conclude with final remarks in Section~\ref{sec:remarks}. 
We include three appendices: Appendix~\ref{app:C} contains complementary remarks referred to in the main text; 
Appendix~\ref{app:ell} collects facts about elliptic functions we need; Appendix~\ref{app:CMR} gives an overview over the elliptic CMR models. 

\section{Elliptic Calogero-Moser-Ruijsenaars models}
\label{sec:CMR} 
We introduce the family of elliptic CMR models and the special functions defined by them. 
For the convenience of the reader, we include Appendix~\ref{app:CMR} describing the classification of CMR models and giving an overview over the elliptic CMR models we discuss in this section. 

To aid our discussion of the physics interpretation of the CMR models, we keep in this section the following three positive parameters in our equations: $\hbar$ (Planck's contant), $m$ (particle mass) and $2\ell$ (size of space). 
These three parameters can be scaled to arbitrary constants by changing the physics units; common choices in the literature are $\hbar=1$, $m=1$ or $1/2$, and $\ell=\pi$, $\pi/2$ or $1$. 
In the relativistic case, there is one parameter $\beta\equiv 1/mc$ in addition to the ones in the non-relativistic case, where $c$ is the vacuum velocity of light. 

\subsection{A-type non-relativistic CMR model}\label{sec:eCS} Another name  for this model  in the literature is $A_{N-1}$ elliptic quantum Calogero-Moser system; we call it {\em elliptic Calogero-Sutherland (eCS) model} in the following. 

For arbitrary fixed variable number $N=2,3,\ldots$, the eCS model can be defined by the differential operator 
\begin{align}\label{HN}
H = -\frac{\hbar^2}{2m} \sum_{i=1}^N\frac{\partial^2}{\partial x_i^2} + \frac{g(g-\hbar)}{m} \sum_{1\leq i<j\leq N} \wp_1(x_i-x_j) 
\end{align} 
acting on suitable complex-valued functions $\psi$ of $x=(x_1,\ldots,x_N)\in[-\ell,\ell]^N$. 
The functions $\psi=\psi(x)$ of interest are solutions of the eigenvalue equation
\begin{align}\label{SchEq}  
H\psi=E\psi, 
\end{align} 
with $E$ the corresponding eigenvalue. 

Our notation is motivated by the following physics interpretation: 
For $g\geq \hbar$ (see Remark~\ref{app:C}.\ref{rem:g}), $H$ in \eqref{HN} is a quantum mechanical Hamiltonian describing an arbitrary number, $N$, of identical particles of mass $m$ moving on a circle of circumference $2\ell$; these particles interact with a repulsive two-body potential given by the Weierstrass elliptic function $\wp_1$,  $x_i\in[-\ell,\ell]$ are the particle positions  ($i=1,\ldots,N$), and $g(g-\hbar)/m$ is the coupling constant (see Remark~\ref{app:C}.\ref{rem:coupling}).  
One can then interpret  \eqref{SchEq} as a time independent Schr\"odinger equation, $\psi$ as an energy eigenfunction, and $E$ as the corresponding energy eigenvalue. 
 
 This interpretation suggests that $H$ defines a self-adjoint operator on the Hilbert space $L^2([-\ell,\ell]^N)$ of square-integrable functions on $[-\ell,\ell]^N$ with the usual scalar product,  
\begin{align}\label{product}  
(\psi,\phi) = \int_{[-\ell,\ell]^N} \psi(x)\overline{\phi(x)}d^Nx 
\end{align} 
where the bar indicates complex conjugation. 
Since we interpret this as a quantum mechanical model of identical particles, we are only interested in wavefunctions $\psi$ such that the probability density $|\psi(x_1,\ldots,x_N)|^2$ is invariant under permutations of the particle positions $x_i$; thus, we can restrict \eqref{SchEq} to the domain 
\begin{align}\label{wedge}  
\ell\geq x_1>x_2>\cdots > x_N >- \ell. 
\end{align} 

The physics interpretation of $H$ in \eqref{HN} above remains true as it stands if we replace $\wp_1(x)$ by any potential $V(x)$ (= $2\ell$-periodic real-valued function of $x\in\R$ which is sufficiently well-behaved); what is special about $V(x)=\wp_1(x)$ is that it defines a quantum integrable model, i.e., there exist commuting differential operators 
\begin{align*} 
H_r= \sum_{i=1}^N\left( -\ii \hbar \frac{\partial}{\partial x_i}\right)^r +\text{lower order terms}
\end{align*} 
for $r=1,\ldots,N$, with $H_2=2mH$ and $H_1=P$ where 
\begin{align}\label{PN}  
P = -\ii\hbar\sum_{i=1}^N \frac{\partial}{\partial x_i}
\end{align} 
is the momentum operator (see e.g. Halln\"as (2024) \cite{hallnas2024} for further details and references).

The $H_r$ for $r=1,\ldots,N$ define commuting formally self-adjoint operators describing identical particles. 
This suggests that there exist common eigenfunctions $\psi_\lambda(x)$ of all these operators which are square integrable, with $\lambda$ some label distinguishing these eigenfunctions.
The special functions defined by the eCS model are these common eigenfunctions $\psi_\lambda(x)$.
 
The eigenfunctions $\psi_\lambda(x)$ of interest are functions of the variables 
\begin{equation}\label{zi}  
z_j=\ee^{\ii\pi x_j/\ell}\quad (j=1,\ldots,N) 
\end{equation} 
determined by $g/\hbar$ (dimensionless coupling parameter) and $p$ (elliptic deformation parameter in \eqref{p}), and they are naturally labeled by integer vectors $\lambda=(\lambda_1,\ldots,\lambda_N)$ such that
\begin{align}\label{partition} 
\lambda_1\geq \lambda_2\geq \cdots \geq \lambda_N\geq 0
\end{align} 
(see Remark~\ref{app:C}.\ref{rem:lambda}).  These eigenfunctions are of the form
\begin{align}\label{eJack1}  
\psi_\lambda(x) = \psi_0(x) P_\lambda(z;g/\hbar,p)
\end{align} 
with
\begin{equation}\label{psi0}  
\psi_0(x) = \prod_{1\leq i<j\leq N} \vartheta_1(x_i-x_j) ^{g/\hbar} 
\end{equation} 
and $P_\lambda$ symmetric functions in the variables $z=(z_1,\ldots,z_N)$.
In the trigonometric case $p=0$, the eigenfunctions are as in \eqref{eJack1} with $P_\lambda(z;g/\hbar,0)$  equal to the Jack polynomials $P_{\lambda}^{(\alpha)}(z_1,\ldots,z_N)$ with parameter $\alpha=\hbar/g$ as defined in the book by Macdonald (1995) \cite{macdonald1995}. 

\subsubsection{Eigenfunctions as orthogonal systems}
There is an important difference between the trigonometric and elliptic cases: In the trigonometric case $p=0$, the function $\psi_0(x)$ \eqref{psi0} is the groundstate, i.e., it is the exact eigenstate of $H$ corresponding to the smallest possible eigenvalue; in the elliptic case $p>0$, $\psi_0(x)$ is {\em not} an exact eigenstate of $H$. However, in both cases, $\psi_0(x)$ defines a weight function as follows: by the ansatz $\psi(x)=\psi_0(x)P(z)$ and $\phi(x)=\psi_0(x)Q(z)$, the scalar product $(\psi,\phi)$ in \eqref{product} is identical with the following scalar product between $P$ and $Q$,  
\begin{align} 
\langle P,Q\rangle' =  \oint \frac{dz_1}{2\pi \ii  z_1}\cdots  \oint \frac{dz_N}{2\pi \ii  z_N}  W(z) P(z)Q(z^{-1}) 
\end{align} 
with the integrations counterclockwise on the circles $|z_j|=1$ in the complex $z_j$-planes, $z^{-1}=(z_1^{-1},\ldots,z_N^{-1})=\bar z$, and $W(z) = \psi_0(x)^2\geq 0$ equal to 
\begin{align}\label{W} 
W(z) = \Bigg(\prod_{\substack{i,j=1\\i\neq j}}^N\theta(z_i/z_j;p)\Bigg)^{g/\hbar} , 
\end{align} 
using the theta function in \eqref{theta} and $\overline{Q(z)}=Q(\overline{z})$  (the latter is not obvious but true). In the trigonometric case, one can define the Jack polynomials as an orthogonal system with respect to this scalar product $\langle\cdot,\cdot\rangle'$ (see Macdonald (1995) \cite{macdonald1995}); it is natural to expect this is true also in the elliptic case (see Remark~\ref{app:C}.\ref{rem:product}). 

For latter reference, we note that for $x$ in the domain in \eqref{wedge}, the functions in \eqref{eJack1} are equal to
\begin{align}\label{eJack} 
\psi_\lambda(x)= W(z)^{1/2}P_\lambda(z;g/\hbar,p). 
\end{align} 

\subsubsection{Lam\'e equation}\label{sec:Lame}  
For $N=2$ and with the ansatz $\psi=\psi(x_1-x_2)$, the eigenvalue equation $H\psi=E\psi$ for the eCS operator in \eqref{HN} is reduced to 
\begin{align}\label{Lame} 
\Big( -\frac{d^2}{d x^2} + g(g-1)\wp_1(x) \Bigr)\psi(x)=E\psi(x)
\end{align} 
with $x=x_1-x_2$, renaming $(g/\hbar,mE/\hbar^2)\to (g,E)$ to simplify notation; this is known as {\em Lam\'e equation}. It is a classic equation in mathematical physics which has been studied since a long time; for example, the last chapter in Whittaker \& Watson (1920) \cite{whittaker1920} is devoted to solutions of this equation. 

We regard the Lam\'e equation (or the $N=2$ eCS model) as the simplest non-trivial example of an elliptic CMR system, and we use it in Sections~\ref{sec:Bethe}--\ref{sec:perturbative} as an example to illustrate three techniques that have been used to construct eigenfunctions of elliptic CMR models. For this reason, it is interesting to note that the Lam\'e equation has other physics interpretations than the one discussed above and, depending on the interpretation, it can define different kinds of special functions: one can change variables in \eqref{Lame} from $x$ to $x+\ii\delta$ and rename $\psi(x+\ii\delta)\to \psi(x)$ to obtain 
\begin{align}\label{Lame1} 
\Big( -\frac{d^2}{d x^2} + g(g-1)\wp_1(x+\ii \delta) \Bigr)\psi(x)=E\psi(x). 
\end{align} 
Mathematically, this is just another form of the Lam\'e equation. However, it has a different physics interpretation: since $\wp_1(x+\ii\delta)$ is real-valued for real $x$, it can also be interpreted as a potential, and this potential is qualitatively different from $\wp_1(x)$ and thus describes different physics. To see this, note that 
\begin{align*} 
\lim_{\ell\to\infty}\wp_1(x+\ii\delta)=  \frac{(\pi/2\delta)^2}{\sinh^2(\pi (x+\ii\delta)/2\delta)} =  -\frac{(\pi/2\delta)^2}{\cosh^2(\pi x/2\delta)}, 
\end{align*} 
which is non-singular and attractive; $\wp_1(x+\ii\delta)$ is a $2\ell$-periodized version of this non-singular attractive potential. 
Similarly as \eqref{Lame}, one can interpret \eqref{Lame1} as a reduced model for the relative coordinate $x=x_1-x_2$ in a system of two identical particles on the circle interacting via the two-body potential $g(g-\hbar)\wp_1(x_1-x_2+\ii\delta)$.  The eigenfunctions of interest to physics for these two quantum mechanical models are different and {\em not} simply related by analytic continuation $x\to x+\ii\delta$ (see Remark~\ref{app:C}.\ref{rem:LameLame1}). 

There is another physics interpretation of \eqref{Lame1}  as quantum mechanical model of a single particle moving on the real line, $x\in\R$, in an external $2\ell$-periodic potential $\wp_1(x+\ii\ell)$. This model  describes an electron on the real line exposed to the potential of a one dimensional crystal lattice. One then can use the Bloch theorem to conclude that \eqref{Lame1} has solutions of the form $\psi(x)=\exp(\ii kx)\phi_n(x;k)$ with $-\pi/2\ell< k\leq \pi/2\ell$, where $\phi_n(x;k)$ are $2\ell$-periodic functions in $x$. The corresponding eigenvalues $E_n(k)$ belong to disjoint intervals known as energy band.  This physics interpretation of \eqref{Lame1} defines special functions $\phi_n(x;k)$ of a different kind than the ones obtained by using the physics interpretation discussed further above.  
 
\subsection{A-type relativistic CMR model}\label{sec:eR} Common names for this model in the literature include elliptic quantum Ruijsenaars-Schneider and $A_{N-1}$ elliptic relativistic quantum Calogero-Moser system; we use the name {\em elliptic Ruijsenaars model}.

This model is a relativistic generalization of the eCS model depending on an additional parameter $\beta=1/mc>0$, with $m$ the particle mass and $c$ the vacuum velocity of light. 
It can be defined  by two commuting analytic difference operators, 
\begin{align}\label{Spm}  
S_{\pm 1} = \sum_{i=1}^N \prod_{\substack{j=1\\j\neq i}}^N\left( \frac{\vartheta_1(x_i-x_j\mp\ii g\beta)}{\vartheta_1(x_i-x_j)}\right)^{1/2}\ee^{\mp \ii \hbar\beta\partial_{x_i}}
 \prod_{\substack{j=1\\j\neq i}}^N \left( \frac{\vartheta_1(x_i-x_j\pm\ii g\beta)}{\vartheta_1(x_i-x_j)} \right)^{1/2}
\end{align} 
acting on symmetric functions $\psi$ of $x=(x_1,\ldots,x_N)$ in a subset of $\C^N$ with certain analyticity properties, where $\exp\left( \mp \ii \hbar\beta \partial_{x_i}\right)$ acts on $\psi(x_1,\ldots,x_N)$ by changing the argument $x_i$ to $x_i\mp \ii \hbar\beta$ without affecting the other variables (see e.g. Halln\"as (2024) \cite{hallnas2024} for further details). 

Ruijsenaars (1987) \cite{ruijsenaars1987} showed that 
\begin{align*} 
H_{\mathrm{rel}}= \frac1{2m\beta^2}(S_1+S_{-1}),\quad P_{\mathrm{rel}}=\frac1{2\beta}(S_1-S_{-1})
\end{align*}  are relativistic generalizations of the Hamiltonian $H$ in \eqref{HN} and momentum operator $P$ in \eqref{PN} in the following sense: (i) they, together with the so-called boost operator $B=-m\sum_{i=1}^N x_i$, satisfy the Poincar\'e algebra in 1+1 spacetime dimensions: 
\begin{align*} 
[H_{\mathrm{rel}},P_{\mathrm{rel}}]=0,\quad [B,H_{\mathrm{rel}}]=\ii\hbar P_{\mathrm{rel}},\quad [B,P_{\mathrm{rel}}]=\ii\hbar H_{\mathrm{rel}}/c^2,
\end{align*}  
(ii) they reduce to $H$ and $P$ in the non-relativistic limit:  
\begin{align*} 
H= \lim_{c\to \infty}(H_{\mathrm{rel}}-Nmc^2),\quad P =\lim_{c\to \infty} P_{\mathrm{rel}}, 
\end{align*} 
where $Nmc^2$ is the rest mass energy of $N$ particles of mass $m$, (iii) they define a quantum integrable system, i.e., there exist $2N$ commuting analytic difference operators $S_{\pm r}$, $r=1,2,\ldots,N$, such that $S_{-r}=S_N^{-1}S_{r}$; see e.g.\ Halln\"as (2024) \cite{hallnas2024}. 

This proposal of Ruijsenaars is intriguing from a physics point of view: according to physics folklore, due to the existence of antiparticles with negative energies, a relativistic quantum theory describing a stable system must be a quantum field theory (with infinitely many degrees of freedom, different from quantum mechanics with finitely many degrees of freedom). At face value, the operators $H_{\mathrm{rel}}$ and $P_{\mathrm{rel}}$ define a relativistic quantum mechanical model providing a counter example to this folklore. 
To offer a resolution of this conundrum, we recall  that Ruijsenaars was led to this model by a 1+1 dimensional quantum field theory which is massive and integrable and, for this reason, can be decomposed into sectors where the number of certain soliton excitations  is preserved --- the Ruijsenaars model is an effective description of quantum solitons in such a sector; see Ruijsenaars (1987)  \cite{ruijsenaars1987}.
It would be interesting to understand this relation between the Ruijsenaars model and quantum field theory in detail.

As in the non-relativistic case, the operators $S_{\pm r}$, $r=1,\ldots,N$, define commuting operators, and their common eigenfunctions $\psi_\lambda(x)$ are the special functions defined by the elliptic Ruijsenaars model. The argument $x$ and label $\lambda=(\lambda_1,\ldots,\lambda_N)$ of these eigenfunctions are the same as in the non-relativistic case, and they are symmetric functions of the form 
\begin{align}\label{eMcD}  
\psi_\lambda(x) = W_{\mathrm{rel}}(z)^{1/2} P_\lambda(z;p,q,t)
\end{align} 
with $p$ in \eqref{p} and  
\begin{align} 
q = \ee^{-\pi\hbar \beta/\ell},\quad t = \ee^{-\pi g \beta/\ell}; 
\end{align} 
the function $W_{\mathrm{rel}}(z)$ plays the role of weight function (which is non-negative), and it is given by 
\begin{align} 
W_{\mathrm{rel}}(z) = \prod_{\substack{i,j=1\\i\neq j}}^N \frac{\Gamma(tz_i/z_j;p,q)}{\Gamma(z_i/z_j;p,q)}
\end{align} 
with the {\em elliptic Gamma function} 
\begin{align}\label{eGamma}  
\Gamma(z;p,q)= \prod_{n,m=0}^\infty \frac{(1-p^{n+1}q^{m+1}/z)}{(1-p^nq^m z)} 
\end{align} satisfying $\Gamma(qz;p,q)=\theta(z;p)\Gamma(z;p,q)$ (see Remark~\ref{app:C}.\ref{rem:eGamma}). 
In the trigonometric case $p=0$, the $P_\lambda(z;q,t,0)$ are equal to the Macdonald polynomials $P_{N,\lambda}(z;q,t)$ --- note that $q,t$ are the usual parameters used by Macdonald (1995) \cite{macdonald1995}. 

It is worth mentioning that the operators $S_{-1}$ and $S_{1}$ in \eqref{Spm} can also be written as
\begin{align*} 
D(z;p,q,t) = \sum_{i=1}^N  \prod_{\substack{j=1\\j\neq i}}^N \frac{\theta(tz_i/z_j;p)}{\theta(z_i/z_j;p)} T_{q,z_i}
\end{align*} 
and $D(z;p,q^{-1},t^{-1})$, respectively, with $T_{q,z_i}$ the operator acting on (suitable) functions $P(z_1,\ldots,z_N)$ by changing $z_i$ to $qz_i$ and leaving the other arguments the same. 
The operators $D_{\pm 1}(z;p,q^{\pm 1},t^{\pm 1})$ are obtained from $S_{\pm 1}$ by a similarly transformation with $W_{\mathrm{rel}}(z)^{1/2}$ and changing the variables from $x_i$ to $z_i$ in \eqref{zi}, i.e., the symmetric functions $P_\lambda(z;p,q,t)$ are common eigenfunctions of $D(z;p,q^{\pm 1},t^{\pm 1})$.

\subsection{BC-type CMR models} 
The $BC_N$ variant of the eCS model is a quantum version of a classical integrable system introduced by Inozemtsev,  and we therefore call it  {\em (quantum) Inotzemtsev model}. It can be defined by the differential operator 
\begin{multline}\label{HBCN} 
H = -\frac{\hbar^2}{2m}\sum_{i=1}^N \frac{\partial^2}{\partial x_i^2} + \sum_{i=1}^N\sum_{\nu=0}^3 \frac{g_\nu(g_\nu-\hbar)}{2m}\wp_1(x_i+\om_\nu) \\  +  \frac{g(g-\hbar)}{m}\sum_{1\leq i<j\leq N}[\wp_1(x_i-x_j)+\wp_1(x_i+x_j)] 
\end{multline} 
with 
\begin{align*} 
\om_0=0,\quad \om_1=\ell,\quad \om_2=\ii\delta,\quad \om_3=-\ell-\ii\delta, 
\end{align*} 
depending on 4 coupling parameters $g_\nu$, $\nu=0,1,2,3$, in addition to $g$. 
If all these coupling parameters are $\geq \hbar$, one can interpret this as a quantum mechanical Hamiltonian describing $N$ particles: the positions $x_i$ of the particles are restricted to $0<x_i<\ell$, which is half the circle with circumference $2\ell$. The particles are exposed to an external potential which is a linear combination of $\wp(x_i+\om_\nu)$ for $\nu=0,1,2,3$; this potential is singular and repulsive at $x_i=0$ and $x_i=\ell$ if $g_0>\hbar$ and $g_1>\hbar$, respectively, confining the particles to the half-circle $0<x_i<\ell$. As in the $A_{N-1}$-case, the terms $\propto \wp_1(x_i-x_j)$ account for two-body interactions between particle $i$ and $j\neq i$, but there are additional non-trivial boundary effects at $x_i=0$ and $x_i=\ell$ taken into account by adding mirror particles at positions $-x_j$ at the other half of the circle: the terms $\propto \wp_1(x_i+x_j)$ describe interactions between a particle at $x_i$ with a mirror particle at $-x_j$ (for $j=i$, such a term is part of the external potential term due to the  well-known identity $\wp_1(2x_i)=(1/4)\sum_{\nu=0}^3\wp_1(x_i+\om_\nu)$). 

In the special case $N=1$, the eigenvalue equation \eqref{SchEq} for the Inozemtsev operator in \eqref{HBCN} can be written as 
\begin{align}\label{Heun} 
\Big( -\frac{\partial^2}{\partial x^2} + \sum_{\nu=0}^3 g_\nu(g_\nu -1)\wp_1(x+\om_\nu) \Big) \psi(x) = E\psi(x)
\end{align} 
for $x=x_1$, renaming $(g_\nu/\hbar,2mE/\hbar^2)\to (g_\nu,E)$ to simplify notation; this is known as {\em Heun equation}. 
Note that the Heun equation  \eqref{Heun} reduces to the Lam\'e equation \eqref{Lame} in the special cases $(g_0,g_1,g_2,g_3)=(g,0,0,0)$ and $(g_0,g_1,g_2,g_3)=(g,g,g,g)$ (the latter case requires also  scaling $(x,E)\to (x/2,4E)$). Moreover, in the trigonometric limit and for $\ell=\pi$, the Heun equation reduces to 
\begin{align*} 
\Bigg(-\frac{\partial^2}{\partial x^2} + \frac{a^2-\tfrac{1}{4}}{\sin^2(x)} +  \frac{b^2-\tfrac{1}{4}}{\cos^2(x)} \Bigg) 
\psi(x) = E\psi(x)\quad 
\end{align*} 
with $a=g_0-\tfrac{1}{2}$ and $b=g_1-\tfrac{1}{2}$, which corresponds to a famous exactly solvable model in quantum mechanics known as the {\em trigonometric P\"oschl-Teller model}. As is well-known, the eigenfunctions of the latter model are given by Jacobi polynomials.  

The trigonometric limit of the Hamiltonian $H$ in \eqref{HBCN} has eigenfunctions given by polynomials defining well-studied many-variable generalizations of the Jacobi polynomials; the Inozemtsev model has eigenfunctions defining elliptic generalizations of these many-variable Jacobi polynomials. 

A quantum integrable $BC_N$-version of the elliptic Ruijsenaars model providing a relativistic generalization of the Inozemtsev model was discovered by van Diejen (1994) \cite{vandiejen1994}. It depends on 8 coupling parameters in addition to the Macdonald parameters $q,t$. The eigenfunctions of this model in the trigonometric case are given by {\em Koornwinder polynomials} (Koornwinder (1992)). Thus, the elliptic van Diejen model defines elliptic generalizations of the Koornwinder polynomials.

\subsection{Non-stationary CMR equations}\label{sec:nonstationary}  
These generalizations of elliptic CMR models are also known as {\em Knizhnik--Zamolodchikov--Bernard (KZB) equations} in the literature. 
We start with the Lam\'e equation as a simple example motivating this generalization, and then describe this generalization in other cases. 

For $g\in\C$, the function $\psi(x)=\sin(x)^g$ of $x\in\C$ is a solution of the differential equation
\begin{align}\label{nrLame} 
\Bigg(-\frac{\partial^2}{\partial x^2} + \frac{g(g-1)}{\sin^2(x)}\Bigg) \psi(x) = E\psi(x)
\end{align} 
(with $E=g^2$). This equation is the trigonometric limit of the Lam\'e equation \eqref{Lame}. 
Thus, one might hope that $\psi(x)=\vartheta_1(x)^g$ satisfies the Lam\'e equation, but this is not the case. 
The reason is that, while $\psi(x)=\sin(x)$ satisfies $-\partial_x^2\psi(x)=E\psi(x)$ (which is \eqref{nrLame} for $g=1$), the corresponding equation satisfied by $\psi(x)=\vartheta_1(x)$ has an additional term proportional to $\partial_\tau\psi(x)$ where 
\begin{align} 
\tau=\frac{\ii\delta}{\ell} 
\end{align} 
 is the half-period ratio; this equation satisfied by the theta function is known as the heat equation; see \eqref{heat}. 
Using this heat equation and $\wp_1(x)=-\partial_x^2\ln\vartheta_1(x)$, one can verify that $\psi(x)=\vartheta_1(x)^g$ is a solution of the following equation, 
\begin{align}\label{nLame} 
\Big(\frac{\ii\pi}{2\ell^2}\kappa \frac{\partial}{\partial\tau} -\frac{\partial^2}{\partial x^2} + g(g-1)\wp_1(x) \Bigr)\psi(x)=E\psi(x)
\end{align} 
for $\kappa=2g$. Eq.\ \eqref{nLame} is known as the {\em non-stationary Lam\'e equation}; one can think of it as a deformation of the Lam\'e equation by a complex parameter $\kappa$. 

The non-stationary Lam\'e equation  is the simplest example of a non-stationary generalization of an elliptic CMR systems. Other such generalizations include the {\em non-stationary eCS equation} given by 
\begin{align}\label{nSchEq} 
\Big( \ii \frac{\hbar^2\pi}{2m\ell^2}\kappa \frac{\partial}{\partial\tau}  +  H\Big)\psi =E\psi
\end{align} 
with $H$ the eCS operator in \eqref{HN}, the {\em non-stationary Inozemtsev equation} in \eqref{nSchEq} with $H$ the Inozemtsev operator in \eqref{HBCN}, and the non-stationary generalizations of the deformed and generalized models discussed in Section~\ref{sec:generalized} which are defined in a similar manner. One would expect that there exist corresponding non-stationary generalizations of the relativistic CMR systems, but this remains to be understood (see Remark~\ref{app:C}.\ref{rem:open}).

A reason for why these non-stationary generalizations are interesting is that they are easier to solve for non-zero $\kappa$ than for $\kappa=0$. 
We discuss in this paper two different kinds of results in this direction: (A) constructions of explicit solutions of non-stationary CMR equations for particular non-zero values of $\kappa$, (B) constructions of solutions $E$ and $\psi$ of conventional elliptic CMR equations, $H\psi=E\psi$, as a limit $\kappa\to 0$ of solutions of the corresponding non-stationary equation in \eqref{nSchEq}. 

As for (A), we gave already one example: $\vartheta_1(x)^g$ is a solution of the non-stationary Lam\'e equation \eqref{nLame} for $\kappa=2g$.  
There are many more known explicit solutions of non-stationary CMR equations; see Section~\ref{sec:nonstationary} for other examples. 
This makes the non-stationary CMR equation interesting from a mathematics point of view.
To give an idea about (B), we note that the non-stationary Lam\'e equation is invariant under the following transformation, 
\begin{align}\label{symm} 
\psi(x)\to \tilde\psi(x)=C\psi(x),\quad E\to \tilde E=E +  \ii \frac{\pi}{\ell^2}\kappa \frac{\partial_\tau C}{C} 
\end{align} 
for arbitrary differentiable non-zero $C=C(\tau)$ (independent of $x$), and the same is true for \eqref{nSchEq} in units where $\hbar=m=1$. 
Thus, the {\em generalized eigenvalue} $E$ in a non-stationary equation can be changed to an arbitrary value by changing the eigenfunction normalization (see Remark~\ref{app:C}.\ref{rem:Egen}). 
This can be used to construct solutions $\psi$ and $E$ of the elliptic CMR eigenvalue equation $H\psi=E\psi$; this is explained in a simple example in Section~\ref{sec:Variant||}.  

\subsection{Deformed elliptic CMR models}\label{sec:deformed} 
As first discovered by Chalykh, Feigin \& Veselov (1998) in a special case, CMR models have integrable deformations describing arbitrary numbers of two kinds of particles; see e.g.\ Sergeev \& Veselov (2004) \cite{sergeev2004}. By-now, such generalizations are known for all (stationary) CMR models (including rational, hyperbolic, trigonometric and elliptic cases). Moreover, in the trigonometric cases, a lot is known about the eigenfunctions of the deformed models. In this section, we discuss deformed elliptic CMR models.

From a quantum mechanics point of view, deformed CMR models have peculiar features, including particles with negative mass, non-existence of a classical limit $\hbar\to 0$, and singular wave functions. However, the deformed CMR models fit into a generalized quantum mechanical framework in the sense that there is a well-defined scalar product promoting the eigenfunctions to an orthogonal basis in a Hilbert space (as discussed in Section~\ref{sec:product}, this has been proved in the A-type trigonometric cases, and we believe it is true in general).
Moreover, many other mathematical results known for conventional CMR models have natural generalizations to the deformed case. 
We believe that the unusual features of deformed CMR models are due to them not describing conventional quantum particles but solitons in integrable quantum field theories, similarly as the relativistic CMR models. While quantum field theory is beyond our scope, we mention it to convey that the deformed CMR models are interesting for mathematics {\em and} physics.

\subsubsection{Deformed eCS model}\label{sec:deformedeCS}  
This model can be written in terms of arbitrary numbers, $N$ and $M$, of  two kinds of variables $x=(x_1,\ldots,x_N)$ and $\tilde x=(\tilde x_1,\ldots,\tilde x_M)$. 
It can be defined by the following operator (Sergeev \& Veselov (2005) \cite{sergeev2005}), 
\begin{align}\label{HNM} 
H_{N,M}(x,\tilde{x};g) =&  -\frac{\hbar^2}{2m} \sum_{i=1}^N\frac{\partial^2}{\partial x_i^2} + \frac{g(g-\hbar)}{m} \sum_{1\leq i<j\leq N} \wp_1(x_i-x_j)\nonumber \\
& +\frac{\hbar g}{2m} \sum_{i=1}^M\frac{\partial^2}{\partial \tilde{x}_i^2} - \frac{\hbar(\hbar^2/g-\hbar)}{m} \sum_{1\leq i<j\leq M} \wp_1(\tilde{x}_i-\tilde{x}_j) \\
& + \frac{\hbar(\hbar-g)}{m} \sum_{i=1}^N\sum_{j=1}^M \wp_1(x_i-\tilde{x}_j)   \nonumber 
\end{align} 
where we find it convenient to write $H_{N,M}(x,y;g)$ instead of the shorter $H$, here and in the rest of this section; in this notation,  the conventional eCS operator in \eqref{HN} is denoted as $H_N(x;g)$. 

Clearly, in the special case $M=0$, the operator in \eqref{HNM}  reduces to the standard eCS operator: $H_{N,0}(x,\cdot;g)=H_N(x;g)$. 
Moreover, the special case $N=0$ corresponds to another standard eCS Hamiltonian: $H_{0,M}(\cdot,\tilde{x};g)=-(g/\hbar)H_M(\tilde{x};\hbar^2/g)$. 
Thus, one can write deformed eCS operator in \eqref{HNM} more compactly as follows, 
\begin{align*} 
H_{N,M}(x,\tilde{x};g)=H_N(x;g)-\frac{g}{\hbar}H_M(\tilde{x};\hbar^2/g) + \frac{\hbar(\hbar-g)}{m} \sum_{i=1}^N\sum_{j=1}^M \wp_1(x_i-\tilde{x}_j) , 
\end{align*} 
which suggests to think of it as Hamiltonian of a system with two kinds of particles interacting with the same two-body interactions given by $\wp_1(x)$; the particles are different since the $x$-particles interact with the coupling parameter $g\geq \hbar$ and the $\tilde x$-particles interact with the coupling parameter $\hbar^2/g$. 
What is peculiar from a conventional quantum mechanics point of view is that the $\tilde x$-particles have a negative mass parameter $-m\hbar/g$. Moreover, the model does not make sense in the limit $\hbar\to 0$, i.e., there is no classical counterpart of the deformed eCS model.

We note that the deformed eCS operator in \eqref{HNM} has the duality property 
\begin{align*} 
H_{N,M}(x,\tilde x;g) = -(g/\hbar)H_{M,N}(\tilde x,x;\hbar^2/g) , 
\end{align*} 
i.e., interchanging the two particles types amounts to swapping 
\begin{align*} 
g/\hbar\leftrightarrow \hbar/g. 
\end{align*}

\subsubsection{Other deformed elliptic CMR models}
 For all stationary CMR models (non-relativistic and relativistic), deformed generalizations are known. For example, the deformed Inotsemtsev models are defined by 
\begin{multline}\label{HBCNM}
H_{N,M}(x,\tilde{x};\{g_\nu\}_{\nu=0}^3,g) = H_N(x;\{g_\nu\}_{\nu=0}^3,g)\\ -\frac{g}{\hbar}H_{M}(\tilde{x};\{\hbar(g+\hbar-2g_\nu)/2g\}_{\nu=0}^3,\hbar^2/g)  \\
+ \frac{\hbar(\hbar-g)}{m}\sum_{i=1}^N\sum_{j=1}^{M}[\wp_1(x_i-\tilde{x}_j)+\wp_1(x_i+\tilde{x}_j)]
\end{multline} 
where $H_N(x;\{g_\nu\}_{\nu=0}^3,g)$ is the Inotsemtsev operator in \eqref{HBCN}. Formulas for the operators defining the deformed elliptic Ruijsenaars and van Diejen models are also known (references can be tracked from Atai (2022) \cite{atai2022}). 

\subsubsection{Eigenfunctions and scalar products}\label{sec:product} 
As discussed in Section~\ref{sec:eCS}, the eigenfunctions of the eCS Hamiltonian in \eqref{HN} are of the form \eqref{eJack1} with $\psi_0(x)\equiv \psi_{0;N}(x;g)$ in \eqref{psi0}. 
In the trigonometric case $p=0$, it is known that the appropriate generalization of $\psi_{0;N}(x;g)$  is 
\begin{align}\label{psi0NM}  
\psi_{0;N,M}(x,y;g) = \frac{\prod_{1\leq i<j\leq N}\vartheta_1(x_i-x_j)^{g/\hbar}\cdot \prod_{1\leq i<j\leq M}\vartheta_1(y_i-y_j)^{\hbar/g}}{\prod_{i=1}^N\prod_{j=1}^M \vartheta_1(x_i-y_j)}
\end{align} 
(with $\vartheta_1(x)\to 2\sin(\pi x/\ell)$), and the generalization of \eqref{eJack1} is 
\begin{align}\label{psi0NMlambda}
\psi_\lambda(x,y;g) = \psi_{0;N,M}(x,y;g)P_\lambda(z,w;g) 
\end{align} 
with $P_\lambda(z,w;g)$ polynomials in the variables $z_j=\exp(\ii\pi x_j/\ell)$, $w_k=\exp(\ii\pi y_k/\ell)$. 
The polynomials $P_\lambda(z,w;g)$ are known as {\em super Jack polynomials}; they are symmetric in the variables $z=(z_1,\ldots,z_N)$ and $w=(w_1,\ldots,w_M)$ separately, and in addition they satisfy the quasi-invariance conditions
\begin{align} 
\left.\left(\hbar \frac{\partial}{\partial z_j} + g \frac{\partial}{\partial w_k} \right)P(z,w)\right|_{z_j=w_k}=0\quad (j=1,\ldots,N;k=1,\ldots,M); 
\end{align} 
see Sergeev \& Veselov (2005) \cite{sergeev2005}. 
Since $\psi_{0;N,M}(x,y;g)$ has singularities, one might think that the deformed eCS model for $p=0$ does not have a quantum mechanical interpretation. However, there exists nevertheless a well-defined scalar product such that the  eigenfunctions $\psi_{0;N,M}(x,y;g)P(z,w;g)$ are orthogonal (Atai, Halln\"as \& Langmann (2019) \cite{atai2019}). Thus, the deformed eCS model for $p=0$ defines a generalized quantum mechanical model. One expects that the eigenfunctions of the deformed eCS model have the form \eqref{psi0NM}--\eqref{psi0NMlambda} even in the elliptic case $p>0$, and this suggests a generalization of the above-mentioned scalar product to $p>0$ ({\em ibid.}).  

Generalizations of these results to the relativistic case are known: the eigenfunctions of the deformed trigonometric Ruijsenaars model are given by {\em super Macdonald polynomials} (Sergeev \& Veselov (2009)), and there exists a scalar product promoting  the super Macdonald polynomials to an orthogonal Hilbert space basis (Atai, Halln\"as \& Langmann (2021)). In the elliptic cases, the construction of eigenfunctions of the deformed relativistic models is an open problem; thus, while there are natural candidates for scalar products, the orthogonality of the pertinent eigenfunctions is open. 

\subsection{Generalized CMR models}\label{sec:generalized} 
We explain further generalizations of CMR models which only exist in the hyperbolic and elliptic cases. 
\subsubsection{Calogero's trick}\label{sec:trick} 
 As explained in Section~\ref{sec:Lame}, by analytic continuation $\wp_1(x)\to\wp_1(x+\ii\delta)$ one can change the Lam\'e equation \eqref{Lame} to another form \eqref{Lame1} with a different physics interpretation. As first observed by Calogero, this idea can be generalized to the eCS model by the following trick: fix $N_1,N_2\in\Z_{\geq 0}$ such that $N=N_1+N_2\geq 1$, and rename \begin{align} 
x_{N_1+j} = y_j-\ii\delta \quad (j=1,\ldots,N_2)  
\end{align} 
in the eCS operator $H$ in \eqref{HN} to obtain  
\begin{align}\label{HN1N2}
H_{N_1,N_2}(x,y;g) = H_{N_1}(x;g)+ H_{N_2}(y;g) +  V_{N_1,N_2}(x,y;g)  
\end{align} 
with $x=(x_1,\ldots,x_{N_1})$ and similarly for $y$, using the shorthand notation 
\begin{align} 
 V_{N_1,N_2}(x,y;g) = \frac{g(g-\hbar)}{m}  \sum_{i=1}^{N_1}\sum_{j=1}^{N_2} \wp_1(x_i-y_j+\ii\delta). 
\end{align} 
Assuming $x\in[-\ell,\ell]^{N_1}$ and $y\in[-\ell,\ell]^{N_2}$, one can interpret the operator in \eqref{HN1N2} as a Hamiltonian describing a system of two kinds of particle with equal mass $m$; while particles of the same kind interact with the singular repulsive two-body potential $\wp(x)$, particles of different types interact with the non-singular attractive potential $\wp(x+\ii\delta)$. The physics described by the quantum mechanical model defined by the Hamiltonian in \eqref{HN1N2} is different from the physics described by the standard eCS Hamiltonian \eqref{HN}, and the corresponding eigenfunctions are different. 

\subsubsection{Generalized CMR models} 
It is easy to extend Calogero's trick above to the deformed eCS operators in \eqref{HNM}. This leads to the following operators describing arbitrary numbers, $(N_1,M_1,N_2,M_2)$, of four particle types, 
\begin{align}
H_{N_1,M_1,N_2,M_2}(x,\tilde x,y,\tilde y;g) = H_{N_1,M_1}(x,\tilde{x};g) + H_{N_2,M_2}(y,\tilde{y};g) 
+ V_{N_1,N_2}(x,y;g) + \nonumber  \\ - (g/\hbar)V_{M_1,M_2}(\tilde x,\tilde y;\hbar^2/g) - (\hbar/g)V_{N_1,M_2}(x,\tilde y;g) - (\hbar/g)V_{N_2,M_1}(\tilde x,y;g) . 
\end{align} For lack of a better name, we propose to call the model defined by this Hamiltonian {\em generalized eCS model}. 
In a similar way, one can define generalized variants of the Inotsemtsev, Ruijsenaars and van Diejen models. 

For each of these four elliptic CMR models, we propose to construct eigenfunctions for the generalized variants of the models, and to try to find a scalar product such that these eigenfunctions are orthogonal. There are natural candidates for these scalar products: they are obtained by applying Calogero's trick to the scalar products of the corresponding deformed models. However, the construction of eigenfunctions is a major project: as for today, most known results are restricted to the special case $(N_1,M_1,N_2,M_2)=(N,0,0,0)$. 

To motivate our proposal, we mention a quantum field theory related to the eCS model which originally was constructed as a means to obtain kernel functions for the eCS model (see Langmann (2000) \cite{langmann2000}). 
It was recently realized that this was only the tip of an iceberg: there is quantum field theory corresponding to a quantum version of a soliton equation which has sectors labeled by the numbers $(N_1,N_2,M_1,M_2)$ of four soliton types, and such a sector corresponds to the generalized eCS model --- the quantum field theory related to the eCS model corresponds to the sectors $(N_1,M_1,N_2,M_2)=(N,0,0,0)$ of this quantum field theory (Berntson, Langmann \& Lenells (2023) \cite{berntson2023}). We thus expect that, to solve this quantum field theory model, one needs the eigenfunctions of the generalized eCS model. In a similar way, we expect that the other generalized CMR models correspond to interesting quantum field theories. 

\section{Solutions by Bethe ansatz}\label{sec:Bethe}
Pioneered by Hans Bethe in 1931 to find the exact eigenvalues and eigenvectors of the Heisenberg spin chain, the term {\em Bethe ansatz} has come to represent a large number of methods for the construction of exact solutions of various types of integrable quantum spin systems, many-body systems and more. A particular class of many-body systems to which it has been applied with great success is the class of elliptic CMR systems. In fact, in the two-body case,
which essentially amounts to the Lam\'e equation \eqref{Lame},
the Bethe ansatz goes back to fundamental work by Hermite, although he of course did not use the term Bethe ansatz.
In this section, we describe Hermite's method and present a brief overview of various multi-variable generalisations obtained in more recent years. 

\subsection{Hermite's solution of the Lam\'e equation}\label{sec:Hermite} 
When $n=-g\in\Z_{\geq 1}$ (see Remark~\ref{app:C}.\ref{rem:g=-n}), the Lam\'e equation \eqref{Lame} has meromorphic solutions $\psi$ with periodicity properties
\begin{equation*}
\psi(x+\ell) = B_\ell \psi(x),\ \ \ \psi(x+\ii\delta) = B_\delta \psi(x)
\end{equation*}
for some (complex) numbers $B_\ell$, $B_\delta$. Following Krichever and Zabrodin (1995), the terminology {\em double-Bloch functions} and {\em Bloch multipliers} is often used for such $\psi$ and $B_\ell$, $B_\delta$, respectively.
It was Hermite who first constructed such solutions of a Bethe ansatz form: one considers an explicit expression for a function depending on $n$ parameters $t_1,\ldots,t_n\in\mathbb{C}$, and whenever they satisfy a set of $n$ so-called Bethe equations the function is a solution.

In the simplest nontrivial case $n=1$, the solutions are of the form
\begin{equation*}
\psi(x;\xi,t) = \ee^{\xi x}\frac{\vartheta_1(x-t)}{\vartheta_1(x)}.
\end{equation*}
We note that 
\begin{equation*}
\frac{1}{\psi}\frac{d\psi}{dx} = \frac{\vartheta_1^\prime(x-t)}{\vartheta_1(x-t)}-\frac{\vartheta_1^\prime(x)}{\vartheta_1(x)}+\xi
\end{equation*}
and consequently that
\begin{equation*}
\frac{d}{dx}\left(\frac{1}{\psi}\frac{d\psi}{dx}\right) = \wp_1(x)-\wp_1(x-t).
\end{equation*}
Since $(\psi^{-1}\psi^\prime)^\prime=\psi^{-1}\psi^{\prime\prime}-(\psi^{-1}\psi^\prime)^2$, it follows that the $n=1$ Lam\'e equation is satisfied by $\psi$ if and only if
\begin{equation*}
\wp_1(x-t)+\wp_1(x)-E = \left(\frac{\vartheta_1^\prime(x-t)}{\vartheta_1(x-t)}-\frac{\vartheta_1^\prime(x)}{\vartheta_1(x)}+\xi\right)^2.
\end{equation*}

The left-hand side (LHS) is a manifestly elliptic function, having only double poles at points congruent to $x=0$ and $x=t$. From the quasi-periodicity properties \eqref{vtheta1qper} of $\vartheta_1(x)$, it is readily inferred that also the right-hand side (RHS) is an elliptic function. Moreover, it clearly has the same leading terms as the LHS in its Laurent expansions around $x=0$ and $x=t$. Hence, if we choose the parameters $\xi,t\in\mathbb{C}$ such that the residues of the RHS vanishes, then Liouville's theorem ensures that $\text{LHS}-\text{RHS}$ equals a constant, which vanishes for an appropriate choice of $E$. More specifically, it is easily seen that vanishing of residues amounts to
\begin{equation*}
\xi = \frac{\vartheta_1^\prime(t)}{\vartheta_1(t)}
\end{equation*}
and, by comparing the LHS and RHS at a suitable point (e.g.~as $x\to 0$), one finds that $E=-\wp_1(t)+\text{constant}$ (see Remark~\ref{app:C}.\ref{rem:constant}).

In the general-$n$ case, one makes the ansatz
\begin{equation}
\psi(x;\xi,t) = \ee^{\xi x}\prod_{j=1}^n \frac{\vartheta_1(x-t_j)}{\vartheta_1(x)},
\end{equation}
with $t=(t_1,\ldots,t_n)$. Proceeding as above, it is straightforward to show that the Lam\'e equation is satisfied as long as
\begin{equation}
n^2\wp_1(x)+\sum_{j=1}^n \wp_1(x-t_j)-E = \left(\sum_{j=1}^n\frac{\vartheta_1^\prime(x-t_j)}{\vartheta_1(x-t_j)}-n\frac{\vartheta_1^\prime(x)}{\vartheta_1(x)}+\xi\right)^2.
\end{equation}
Requiring the residue of the RHS at $x=0$ to vanish yields
\begin{equation}\label{xieq}
\xi = \sum_{j=1}^n\frac{\vartheta_1^\prime(t_j)}{\vartheta_1(t_j)}
\end{equation}
and vanishing of the residue at $x=t_j$ amounts to the Bethe equation
\begin{equation}\label{Betheeqs} 
\sum_{\substack{k=1\\k\neq j}}^n\left(\frac{\vartheta_1^\prime(t_j-t_k)}{\vartheta_1(t_j-t_k)}-\frac{\vartheta_1^\prime(t_j)}{\vartheta_1(t_j)}+\frac{\vartheta_1^\prime(t_k)}{\vartheta_1(t_k)}\right) = 0\quad (j=1,\ldots,n),
\end{equation}
and similarly as for $n=1$ one finds $E=-(2n-1)\sum_{j=1}^n\wp_1(t_j)+\mathrm{constant}$. 
For generic $E$, the solution space of the Lam\'e equation \eqref{Lame} with $n=-g\in\Z_{\geq 1}$ is spanned by the functions $\psi(\pm x)$.
The only exceptions are the eigenvalues $E$ for which the equation has a doubly-periodic solution, as discussed in detail by Whittaker \& Watson (1920) \cite{whittaker1920}.

\subsection{Generalizations}\label{sec:Bethe_gen}
For $n=-g\in\Z_{\geq 1}$, the non-stationary Lam\'e equation \eqref{nLame} can be viewed as a particular instance of a KZB equation for a function taking values in the zero-weight subspace of an irreducible representation of $\mathfrak{sl}_2$; see e.g.~Varchenko (2003) \cite{varchenko2003}. By computing the $\kappa\to 0$ asymptotics of an integral representation of solutions of this KZB equation, Etingof \& Kirillov (1994) \cite{etingof1994} gave a new derivation of Hermite's solutions of the Lam\'e equation; see Section~\ref{sec:integrals} for details.

Shortly thereafter, their approach was generalised by Felder and Varchenko (1995,1997) \cite{felder1995,felder1997} to the setting of an arbitrary simple Lie algebra $\mathfrak{g}$, thus obtaining meromorphic eigenfunctions of Bethe ansatz type for generalized eCS operators. More specifically, consider a non-degenerate ad-invariant bilinear form $(\cdot,\cdot)$ on $\mathfrak{g}$, a Cartan subalgebra $\mathfrak{h}$ and a set of roots $R\subset \mathfrak{h}^*$; and, to each $\alpha\in R$, associate an element $e_\alpha\in\mathfrak{g}_\alpha$, normalized such that $(e_\alpha,e_{-\alpha})=1$. Then, the relevant Schr\"odinger operator is given by
\begin{equation*}
H_R = -\Delta + \sum_{\alpha\in R} \wp_1((\alpha,x))e_\alpha e_{-\alpha},
\end{equation*}
acting on functions on $\mathfrak{h}$ that take values in the zero-weight subspace of a highest weight representation $U$ of $\mathfrak{g}$. To recover the $A_{N-1}$ operator \eqref{HN} (with $n=-g\in\Z_{\geq 1}$), one should take $\mathfrak{g}=\mathfrak{sl}_N$ and let $U$ be the $nN$-th symmetric power of its defining representation $\mathbb{C}^N$.

We also note that Chalykh, Etingof \& Oblomkov (2003) \cite{chalykh2003} established a different procedure for constructing double-Bloch eigenfunctions of a class of (scalar) elliptic multidimensional Schr\"odinger operators, with the eCS operator \eqref{HN} and its root system generalisations included as special cases; and that Takemura (2003) \cite{takemura2003} obtained meromorphic eigenfunctions of the $N=1$ instance of the Inozemstev operator \eqref{HBCN} by a Bethe ansatz method.

In the relativistic case, Krichever \& Zabrodin (1995) \cite{krichever1995}, Felder \& Varchenko (1996) \cite{felder1996} and Ruijsenaars (1999) \cite{ruijsenaars1999} obtained eigenfunctions of Bethe ansatz type
for a difference Lam\'e operator, essentially equivalent to the $N=2$ Ruijsenaars model.
Chalykh (2007) \cite{chalykh2007} similarly handled the $\text{BC}_1$ Ruijsenaars model, given by a three-term difference operator depending on $8$ parameters and containing the $N=1$ Inozemstev operator \eqref{HBCN} as a limiting case. Billey (1998) \cite{billey1998} generalized the eigenfunction construction used by Felder and Varchenko to the arbitrary-$N$ Ruijsenaars model, with representation theory of the elliptic quantum group associated to $\mathfrak{gl}_N$ playing a central role.

\section{Solutions by integrals}\label{sec:integrals}
As indicated above, Bethe ansatz solutions of the Lam\'e equation \eqref{Lame} with $n=-g\in\mathbb{Z}_{\geq 1}$ can be tied in with solutions by integrals of its non-stationary counterpart \eqref{nLame}. Specifically, from the perspective of conformal field theory and KZB-equations, Etingof \& Kirillov (1994) \cite{etingof1994} obtained solutions of the latter equation that can be written as
\begin{multline}\label{V}
F_{\xi,\kappa}(x;\tau)  = \ee^{\xi x} 
\int_{\Delta_n} dt_1\cdots dt_n \\ \times \prod_{j=1}^n \frac{\ee^{t_j(8\xi/\kappa-2\pi\ii/\ell)}}{\vartheta_1(t_j)^{8n/\kappa}} 
\prod_{1\leq j<k\leq n}\vartheta_1(t_j-t_k)^{8/\kappa} 
\prod_{j=1}^n \frac{\vartheta_1(x-t_j)}{\vartheta_1(x)\vartheta_1(t_j)}
\end{multline}
for suitable integration cycles $\Delta_n\subset\mathbb{C}^n$ (see Remark~\ref{app:C}.\ref{rem:EK}).
In the simplest nontrivial case $n=1$, one can use a ``figure-eight'' contour that encircles $2\ell$ in the clockwise direction and $0$ in the counterclockwise direction.
In the limit $\kappa\to 0$, the integral in \eqref{V} can be computed using a saddle point evaluation, and one obtains the Bethe ansatz solution in Section~\ref{sec:Hermite} 
(see Remark~\ref{app:C}.\ref{rem:saddle}). Far reaching multivariable generalisation of such results to eCS type models were obtained by Felder \& Varchenko; see Section~\ref{sec:Bethe_gen}.

In the remainder of this section, we focus on a different method for obtaining explicit solutions by integrals of elliptic CMR models, namely the {\em kernel function method}. Kernel functions are explicitly known functions that satisfy certain functional identities involving a pair of CMR operators of the same kind (but where parameters can be different); see \eqref{KFid}. It is interesting to note that kernel functions sometimes are correlation functions in some quantum field theory, but it is possible to ignore this: one can write them down and prove their key properties by direct computations. Using simple concrete examples, we explain what (generalized) kernel functions of elliptic CMR systems are, how they can be used to construct eigenfunctions, and we also mention more general such results in the literature. To simplify notation, we set $\hbar=m=1$ throughout this section. 
 
\subsection{Kernel functions}\label{sec:kernel} 
A {\em generalized kernel function} of a pair $(H(x),\tilde{H}(y))$ of non-relativistic elliptic CMR operators $H(x)$ and $\tilde{H}(y)$ (acting on two sets of variables $x$ and $y$ and suppressing the dependence on the half-period ratio $\tau$) is an explicitly known  function $K(x,y)$ obeying the {\em generalized kernel identity}
\begin{align}\label{KFid}  
\Big(\frac{\ii\pi}{2\ell^2}\kappa\frac{\partial}{\partial\tau}   + H(x)-\tilde{H}(y)-C\Big)K(x,y)=0 
\end{align} 
for some constants $\kappa$ and $C$; if \eqref{KFid} holds true for $\kappa=0$, then $K(x,y)$ is called {\em kernel function} and \eqref{KFid} is called {\em kernel identity}. 
Kernel functions for relativistic elliptic CMR models are defined in exactly the same way as in the non-relativistic case; it would be interesting to find the definition and examples of generalized kernel functions for relativistic elliptic CMR systems --- this is an open question. 

\subsubsection{Example} For $N,M\in\Z_{\geq 0}$ such that $N+M>0$, let 
\begin{align}\label{KNM}
K_{N,M}(x,y;g) = \frac{\prod_{1\leq i<j\leq N}\vartheta_1(x_i-x_j)^g\cdot \prod_{1\leq i<j\leq M}\vartheta_1(y_i-y_j)^g}{\prod_{i=1}^N\prod_{j=0}^M\vartheta_1(x_i-y_j)^g}
\end{align} 
where $x=(x_1,\ldots,x_N)$ and $y=(y_1,\ldots,y_M)$. 
One can prove that this function obeys the identity  
\begin{align}\label{KNMid}  
\Big( \frac{\ii\pi}{2\ell^2}g(N-M)\frac{\partial}{\partial\tau}  + H_N(x;g)-H_M(y;g)-C_{N,M}   \Bigr)K_{N,M}(x,y;g)=0 
\end{align}  
for $H_N(x;g)$ the eCS operator in \eqref{HN} and similarly for $H_M(y;g)$, with $C_{N,M}$ a known constant proportional to $N-M$; see Langmann (2010) \cite{langmann2010}. Thus, the function $K_{N,M}(x,y;g)$ in \eqref{KNM} is a kernel function of  the pair $(H_N(x;g),H_M(y;g))$ of eCS operators in the special cases $M=N$, and in all other cases it is a generalized kernel function of this pair with $\kappa=(N-M)g$ (see Remark~\ref{app:C}.\ref{rem:neCS}). 

\subsection{Kernel function method}\label{sec:kernelmethod}  
The kernel function methods allows to construct a solution $\psi(x)$ of the non-stationary CMR equation 
\begin{align}\label{nellCMR} 
\Big(\frac{\ii\pi}{2\ell^2}\kappa\frac{\partial}{\partial\tau}   + H(x)\Big)\psi(x)=E\psi(x)
\end{align} 
from a solution $\tilde\psi(y)$ of 
\begin{equation}\label{nCMRy} 
\Big(\frac{\ii\pi}{2\ell^2}\kappa\frac{\partial}{\partial\tau}   + \tilde H(y)\Big)\tilde \psi(y)=\tilde E\tilde\psi(y)
\end{equation}  
using a generalized kernel function $K(x,y)$ of the pair of operators $(H(x),\tilde{H}(y))$ with the same $\kappa$.   
We stress that the kernel function method explained below works for arbitrary $\kappa$, including $\kappa=0$ --- in the latter case, the method provides an eigenfunction of $H(x)$ from an eigenfunction of $\tilde{H}(y)$.

\subsubsection{Kernel function transform}\label{sec:strategy}  
The idea is to start from a solution  $\tilde \psi(y)$ of the non-stationary equation \eqref{nCMRy} and to use the generalized kernel function $K(x,y)$  to construct a solution of \eqref{nellCMR} by integrating the product $K(x,y)\tilde\psi(y)$ over some suitable integration domain $\cC$: 
\begin{equation}\label{psi(x)} 
\psi(x)\equiv \int_{\cC} K(x,y)\tilde\psi(y)dy . 
\end{equation} 
To see how this works, use \eqref{KFid} and the product rule for the $\tau$-derivative to compute 
\begin{align*} 
\Big(\frac{\ii\pi}{2\ell^2}\kappa\frac{\partial}{\partial\tau}   + H(x)-C\Big)K(x,y)\tilde\psi(y) 
= \tilde \psi(y)\tilde H(y)K(x,y)  -K(x,y) \tilde H(y)\tilde\psi(y) \nonumber 
\\ + K(x,y)\underbrace{\Big(\frac{\ii\pi}{2\ell^2}\kappa\frac{\partial}{\partial\tau} + \tilde H(y)\Big)\tilde\psi(y)}_{\tilde EK(x,y)\tilde\psi(y)}, 
\end{align*} 
subtracting and adding the same term $K(x,y) \tilde H(y)\tilde\psi(y) $. Recall that, in all our examples, $H(y)$ is a sum of terms proportional to second derivatives $\partial_{y_i}^2$ plus potential terms and, for this reason, $\tilde \psi(y)\tilde H(y)K(x,y)  -K(x,y) \tilde H(y)\tilde\psi(y)$ is a linear combination of total derivative terms. 
Thus, $\psi(x)$ in \eqref{psi(x)} solves \eqref{nellCMR} provided the integration domain $\cC$ is such that, (i) the integral in \eqref{psi(x)} is well-defined, (ii) the pertinent linear combination of integrals of boundary terms 
\begin{align}\label{BT} 
\int_{\cC} \frac{\partial}{\partial y_i}\Big( \tilde \psi(y)\frac{\partial}{\partial y_i}K(x,y)  -K(x,y) \frac{\partial}{\partial y_i} \tilde\psi(y) \Big)dy 
\end{align} 
vanishes. 

\subsubsection{Example:}\label{sec:example} 
We derive a solution of the $N=2$ non-stationary eCS equation, \eqref{nellCMR} for $H(x)=H_2(x;g)$, $\kappa=g$, and $x=(x_1,x_2)$, 
using \eqref{KNMid} for $(N,M)=(2,1)$. 

Since  $\tilde H(y)=-(1/2)\partial_y^2$, it is easy to find solutions of \eqref{nCMRy}: $\tilde\psi(y)=\exp(\ii kx)$ with $\tilde E=k^2$ ($k\in\R$), and $K(x,y)\tilde\psi(y)$ is equal to
 \begin{align*} 
 \frac{\vartheta_1(x_1-x_2)^g}{\vartheta_1(x_1-y)^g\vartheta_1(x_2-y)^g}\ee^{\ii k y}
 = \frac{\ee^{\ii\pi g(x_1+x_2)/\ell}}{\ii^{2g}}\frac{\vartheta_1(x_1-x_2)^g\xi^{\ell k/\pi+g}}{\theta(z_1/\xi;p)^g \theta(z_2/\xi;p)^g}, 
 \end{align*} 
 changing variables to $z_j=\exp(\ii \pi x_j/\ell)$, $\xi=\exp(\ii \pi y/\ell)$ and using the relation between $\vartheta_1(x)$ and $\theta(z;p)$ spelled out in the introduction. 
 Thus, if we choose $\ell k/\pi+g\equiv \lambda$ to be an integer and drop an irrelevant multiplicative constant, then we can choose the integration contour $\cC$ as the straight line from $-\ell-\ii\epsilon$ to $+\ell-\ii\epsilon$ in the complex $y$-plane (for suitable $\epsilon>0$) to get a solution
 \begin{align*} 
 \psi(x) = \ee^{\ii k'(x_1+x_2)}\vartheta_1(x_1-x_2)^g  \oint \frac{d\xi}{2\pi\ii\xi}\frac{\xi^{\lambda}}{\theta(z_1/\xi;p)^g\theta(z_2/\xi;p)^g}, 
 \end{align*} 
 for $k'=\pi g/\ell$ (see Remark~\ref{app:C}.\ref{rem:contour} for further details).
  
 We observe that, due to translational invariance, we can multiply an eigenfunction $\psi(x)$ of $H_2(x;g)$ by a factor $\exp(\ii p(x_1+x_2))$ to get another eigenfunction, for arbitrary $p\in\C$. Thus, by choosing $p$ such that $p+g\pi/\ell=\lambda_2$ and renaming $\lambda=\lambda_1-\lambda_2$ where $\lambda_1,\lambda_2$ are integers such that $\lambda_1-\lambda_2\geq 0$, we obtain a solution of the $N=2$ non-stationary eCS equation for $\kappa=g$ of the form 
\begin{align}\label{psilambda} 
\psi_\lambda(x)=\psi_0(x)P_\lambda(z) 
\end{align} 
where $\psi_0(x)=\vartheta_1(x_1-y)^g$ and 
\begin{align} 
P_\lambda(z) = (z_1z_2)^{\lambda_2}  \oint \frac{d\xi}{2\pi\ii\xi}\frac{\xi^{\lambda_2-\lambda_1}}{\theta(z_1/\xi;p)^g\theta(z_2/\xi;p)^g}\quad (N=2).
\end{align}

\subsection{Generalizations} The kernel functions $K_{N,N}(x,y;g)$ given in Section~\ref{sec:kernel} appeared first in the perturbative solution of the eCS model discussed in Section~\ref{sec:eCSsolution}   (Langmann (2000) \cite{langmann2000}). The generalized kernel functions in \eqref{KNM} are a special case of generalized kernel functions $K_{N,\tilde{N},M,\tilde{M}}(x,\tilde{x},y,\tilde{y};g)$ for pairs of deformed eCS operators as in \eqref{HNM} obtained by a {\em source identity} (Langmann (2010) \cite{langmann2010}). The source identity technique is simple and powerful: it allowed to obtain generalized kernel function identities for pairs of deformed Inozemtsev operators as in \eqref{HBCNM} (Langmann \& Takemura (2012)) and, in the relativistic cases, it led to the kernel functions (only cases with $\kappa=0$) of deformed Ruijsenaars operators (Halln\"as, Langmann \& Atai (2014)) and van Diejen operators (Atai (2022) \cite{atai2022}; see the latter paper for further references). We mention that the kernel functions for pairs of Ruijsenaars and van Diejen operators were first obtained by Ruijsenaars (2004) and Komori, Noumi \& Shiraishi (2009), respectively, and in some cases the kernel function identities are proved not only for the Hamiltonian but the whole family of commuting operators; see Atai (2022)  \cite{atai2022} for references. 

The procedure in Section~\ref{sec:example} can be continued to $N=3,4,\ldots$: By induction, using the kernel function in \eqref{KNM} for $(N,M)=(N,N-1)$ and translation invariance (see Remark~\ref{app:C}.\ref{rem:p}), one can construct solutions $\psi_\lambda(x)$ of the non-stationary eCS equation  as in \eqref{psilambda} with $P_\lambda(z)$ elliptic generalizations of the Jack polynomials given by an explicit $N(N-1)/2$-fold integral depending on integer vectors $\lambda=(\lambda_1,\ldots,\lambda_N)$ satisfying \eqref{partition} (Atai \& Langmann (2020) \cite{atai2020}). 

In our examples above we gave integral representations of solutions of a non-stationary Lam\'e equation ($\kappa\neq 0$). We mention that, using kernel functions for pairs of Inozemtsev operators, one can obtain integral representations of the Lam\'e and Heun equations ($\kappa=0$) using the kernel function method; see e.g.\ Atai \& Langmann (2018) \cite{atai2018} for examples. 

Atai \& Noumi (2023) \cite{atai2023} obtained integral representations of certain eigenfunctions of van Diejen operators using the kernel function method; other such results in the literature can be tracked from this reference. 
 
\section{Perturbative solutions}\label{sec:perturbative} 
One can construct eigenfunctions of elliptic CMR models by expanding in the elliptic deformation parameter $p$. 
In this section, we explain two variants of this method in a simple example. We also shortly discuss the literature on this subject. 

\subsection{Perturbative solutions of eCS model}\label{sec:eCSsolution}  
We consider the non-stationary eCS equation \eqref{nSchEq} with $H=H_2(x;g)$ in \eqref{HN} for $N=2$. 

\subsubsection{Formal perturbation theory}\label{sec:P1}
The Weierstrass $\wp$-function can be expanded as 
\begin{align} 
\wp_1(x)  = -\Big(\frac{\pi}{\ell}\Big)^2 \sum_{m=1}^\infty  m\Big( z^m  +  \sum_{\nu=1}^\infty p^{m\nu}(z^m+z^{-m}) \Big)
\end{align} 
for $z=\exp(\ii\pi x/\ell)$ such that $p<|z|<1$ (see \eqref{wpseries} {\em ff}). This allows to write the non-stationary eCS equation for $N=2$ as $L\psi=0$ with  
\begin{multline*} 
L\equiv  
-\kappa p\partial_p + \tfrac{1}{2}(z_1\partial_{z_1})^2 + \tfrac{1}{2}(z_2\partial_{z_2})^2 -\cE \\ 
-\gamma \sum_{m=1}^\infty  m\Big((z_1/z_2)^m+ \sum_{\nu=1}^\infty p^{m\nu}\big((z_1/z_2)^m  +  (z_1/z_2)^{-m}\big) \Big) , 
\end{multline*} 
where $\cE\equiv (\ell/\pi)^2E$ and $\gamma\equiv g(g-1)$; we changed variables to $z_j=\exp(\ii\pi x_j/\ell)$ and dropped an overall factor $(\pi/\ell)^2$. 
Thus, the action of $L$ on a plane wave function $\exp(\ii(p_1x_1+p_2x_2)=z_1^{s_1}z_2^{s_2}$, $s_i=p_i\ell/\pi\in\R$, is a linear superposition of functions 
\begin{align} 
f_{n,k}(z;s) \equiv z_1^{n+s_1}z_2^{-n+s_2}p^k
\end{align} 
for $n\in\Z$, $k\in\Z_{\geq 0}$; more specifically, 
\begin{align*} 
Lf_{n,k} = (\cE^{(n)}-\cE-k\kappa) f_{n,k} -
\gamma\sum_{m=1}^\infty m\big( f_{n+m,k} + \sum_{\nu=1}^\infty (f_{n+m,k+\nu m}+f_{n-m,k+\nu m} ) \big)
\end{align*} 
with $\cE^{(n)}\equiv \frac12(n+s_1)^2+\frac12(n-s_1)^2$. 
This action is triangular: $Lf_{n,k}$ is a linear combination of functions $f_{n',k'}$ with $(n',k')>(n,k)$ --- by the latter we mean that either (i) $k'>k$ or (ii) $k=k'$ and $n'>n$. Thus, $Lf=0$ has a solution of the form 
\begin{align} 
f(z;s) = \sum_{(n,k)\geq (0,0)}a_{n,k} f_{n,k},\quad \cE=\sum_{k=0}^\infty \cE_k p^k, 
\end{align} 
where all coefficients $a_{n,k}$ can be computed recursively as linear combinations of $a_{n',k'}$ with $(n',k')<(n,k)$, starting from $a_{0,0}=1$ (the latter is a convenient choice). 
To find these recursions, we use the formula for $Lf_{n,k}$ above and change summation indices to compute $A_{k,n}$ such that  $Lf= \sum_{(n,k)\geq (0,0)}A_{n,k} f_{n,k}$; 
solving $Lf=0$ by imposing the conditions $A_{n,k}=0$ we find 
\begin{multline}\label{Ank}  
(\cE^{(n)}-\cE_0-k\kappa)a_{n,k} =  \sum_{k'=1}^{k} \cE_{k'}a_{n,k-k'}
\\ + \gamma \sum_{m=1}^n m a_{n-m,k}+ \gamma  \sum_{\nu=1}^k \sum_{m=1}^{\lfloor\frac{k}{\nu}\rfloor}m\big(a_{n-m,k-\nu m} + a_{n+m,k-\nu m} \big). 
\end{multline} 
For $(n,k)=(0,0)$ this reduces to $(\cE^{(0)}-\cE_0)a_{0,0}=0$, which implies (since $a_{0,0}=1$)  
\begin{align} 
\cE_0=\cE^{(0)}=\frac12(s_1^2+s_2^2),\quad \cE^{(n)}-\cE_0 = n(n+s_1-s_2). 
\end{align} 
One can verify that the recursion relations in \eqref{Ank} can be solved with the constraint 
\begin{align}\label{acond1}
a_{n,k}=0\quad (n<-k). 
\end{align}
\subsubsection{Variant I}\label{sec:VariantI}\label{sec:P2}
To proceed, we set $\kappa=0$. Then there are cases where the recursion relations in \eqref{Ank} need to be treated in a different way from the others: for $n=0$ and $k>0$, the factor multiplying $a_{0,k}$ is zero, which means that $a_{0,k}$ can be arbitrary; instead, \eqref{Ank} determines $\cE_k$ --- this is the mechanism leading to non-trivial eigenvalues for $\kappa=0$. While $a_{0,k}$ can be chosen arbitrarily, a minimalistic choice is  (see Remark~\ref{app:C}.\ref{rem:minimalistic})
\begin{align}\label{acond2}  
a_{0,k}=0\quad (k>1). 
\end{align}
In all other cases, $a_{n,k}$ is determined recursively and in a unique manner by \eqref{Ank}. Thus, by  imposing the conditions $\kappa=0$, \eqref{acond1} and \eqref{acond2}, the recursion relations in \eqref{Ank} have a unique solution $a_{n,k}=a_{n,k}^{(I)}(s)$ and $\cE_k=\cE_k^{(I)}(s)$ (see Remark~\ref{app:C}.\ref{rem:resonance}), and this determines eigenfunctions $f^{(I)}(z;s)$ and corresponding eigenvalues $\cE^{(I)}(s)$: 
\begin{align}\label{fI} 
f^{(I)}(z;s) = \sum_{k=0}^\infty \sum_{n=-k}^\infty a_{n,k}^{(I)}(s)z_1^{s_1+n}z_2^{s_2-n}p^k,\quad \cE^{(I)}(s)=\sum_{k=0}^\infty \cE_k^{(I)}(s)p^k. 
\end{align} 

\subsubsection{Kernel function transform}\label{sec:P3}
The eigenfunctions $f^{(I)}(z;s)$ of the eCS operator $H_2(x;g)$ constructed above are infinite series which, at best, converge in the region $p<|z_1/z_2|<1$. 
From a quantum mechanics point of view, such eigenfunctions are not interesting. 
However, one can use the kernel function $K_{N,M}(x,y;g)$ in \eqref{KNM} for $N=M=2$ to transform $f^{(I)}(z;s)$ to eigenfunctions of interest in physics: as explained in Section~\ref{sec:kernel}, since $(H_2(x;g)-H_2(y;g))K_{2,2}(x,y;g)=0$ and $H_2(y;g)f^{(I)}(\xi;s)=E(s)f^{(I)}(\xi;s)$ for $\xi_j=\exp(\ii\pi y_j/\ell)$ and $E(s)\equiv (\pi/\ell)^2\cE^{(I)}(s)$, the function 
\begin{align*} 
\psi(x) = \mathrm{const.} \int_{\cC} dy_1dy_2 \, K_{2,2}(x,y;g) f^{(I)}(\xi;s)
\end{align*} satisfies $H_2(x;g)\psi(x) = E(s)\psi(x)$ provided certain technical conditions are fulfilled --- as will be seen, these technical conditions are important.

It turns out that, to find an appropriate integration contour $\cC$, it is enough to investigate the integral transform of the function $\xi_1^{s_1}\xi_2^{s_2}$. 
Changing variables and using the relation between  $\vartheta_1(x)$ and $\theta(z;p)$ stated in the introduction to write $K_{2,2}(x,y;g)\xi_1^{s_1}\xi_2^{s_2}$  as (we dropped an irrelevant factor --- see Remark~\ref{app:C}.\ref{rem:K2})
\begin{align*} 
\underbrace{\vartheta_1(x_1-x_2)^g}_{\psi_0(x)}  \xi_1^{s_1-g/2}\xi_2^{s_2+g/2}\frac{\theta(\xi_1/\xi_2;p)^g}{\prod_{i,j=1,2}\theta(z_i/\xi_j;p)^g}.
\end{align*} 
Thus, if we choose $s_1=\lambda_1+g/2$, $s_2=\lambda_2-g/2$ for integers $\lambda_1$, $\lambda_2$, then we can integrate $K_{2,2}(x,y)\xi_1^{s_1}\xi_2^{s_2}$ over $y_j$ along straight lines from $-\ell-\ii\epsilon_j$ to $+\ell-\ii\epsilon_j$ for suitable $\epsilon_j>0$ ($j=1,2$), and this leads to the function $\psi_0(x)F_{\lambda}(z)$ with  
\begin{align} 
F_{\lambda}(z) = \oint_{|\xi_1|=R_1} \frac{d\xi_1}{2\pi\xi_1} \oint_{|\xi_2|=R_2} \frac{d\xi_2}{2\pi\xi_2} \xi_1^{\lambda_1}  \xi_2^{\lambda_2}  \frac{\theta(\xi_1/\xi_2;p)^g}{\prod_{i,j=1}^2\theta(z_i/\xi_j;p)^g}. 
\end{align} 
With this choice, the integration contours are closed and, for this term, boundary terms like in \eqref{BT} vanish. 
At closer inspections, one finds that one should choose $1<R_1<R_2<1/p$ and $\lambda_1\geq \lambda_2\geq 0$ (see Remark~\ref{app:C}.\ref{rem:R1R2}).

With $s$ and the integration contour $\cC$ fixed,  the arguments above generalize to other plane waves $\xi_1^{\lambda_1+n+g/2}\xi_2^{\lambda_2-n-g/2}$ ($n\in\Z$) appearing in the eigenfunction $f^{(I)}(\xi;s)$: they are transformed to $\psi_0(x)F_{\lambda_1+n,\lambda_2-n}(z)$ by this integral transform. This suggests that the integral transform with the kernel function $K_{2,2}(x,y;g)$ transforms $f^{(I)}(z;\lambda_1+g/2,\lambda_2-g/2)$ to the function $\psi_\lambda(x)=\psi_0(x)P_\lambda(x)$ with  
\begin{equation}
P_\lambda(z) = \sum_{k=0}^\infty \sum_{n=-k}^\infty a^{(I)}_{n,k}(\lambda_1+g/2,\lambda_2-g/2)F_{\lambda_1+n,\lambda_2-n}(z) p^k . 
\end{equation} 
One can prove that  $\psi_\lambda(x)$ is a well-defined eigenfunction of $H_2(x;g)$ with corresponding eigenvalues $E_\lambda = (\pi/\ell)^2\cE^{(I)}(\lambda_1+g/2,\lambda_2-g/2)$, as our argument above suggests (Langmann (2014)).

\subsubsection{Variant II}\label{sec:Variant||}\label{sec:P4}
We now explain an alternative way to construct the eigenfunctions $f^{(I)}(z;s)$ of $H_2(x;g)$.  
To solve \eqref{Ank}, we now assume that $\kappa$ is non-zero and such that $n(n+s_1-s_2)-k\kappa$ is always non-zero unless $(n,k)=(0,0)$; this is true if $\kappa$ has a non-zero imaginary part (e.g.). 
Now the cases $n=0$, $k>0$ are no longer special, and we can solve \eqref{Ank} with the constraint 
\begin{align}\label{Econd}  
\cE_k = 0\quad (k\geq 1), 
\end{align}
i.e., we can fix the generalized eigenvalue $\cE$ to be equal to the eigenvalue $\cE_0$ in the trigonometric case. 
This is a huge simplification, leading to much simpler formulas for $a_{n,k}$ as compared to the case $\kappa=0$. 

Imposing the conditions \eqref{acond1} and \eqref{Econd}, the recursion relations in \eqref{Ank} have a unique solution $a_{n,k}=a_{n,k}^{(II)}(s,\kappa)$, and this determines eigenfunctions $f^{(II)}(z;s,\kappa)$ and corresponding eigenvalues $\cE^{(II)}(s)$: 
\begin{align*}
f^{(II)}(z;s,\kappa) = \sum_{k=0}^\infty \sum_{n=-k}^\infty a_{n,k}^{(II)}(s,\kappa)z_1^{s_1+n}z_2^{s_2-n}p^k,\quad \cE^{(II)}(s)=\cE_0(s). 
\end{align*}

We now observe that, while the constant part of $f^{(I)}(z;s)$ (i.e., the coefficient of $z_1^{s_1}z_2^{s_2}$) is $C^{(I)}=a_{0,0}=1$, the constant part of $f^{(II)}(z;s,\kappa)$ is 
\begin{align*}
C^{(II)} = 1+\sum_{k=1}^\infty a^{(II)}_{0,k}(s,\kappa)p^k=C^{(II)}(s,\kappa). 
\end{align*}
Since the constant part of $ f^{(II)}(z;s,\kappa)/C^{(II)}(s,\kappa)$ is 1, it converges to $f^{(I)}(z;s)$ in the limit $\kappa\to 0$  (see Remark~\ref{app:C}.\ref{rem:gauge}). Using \eqref{symm} explained in Section~\ref{sec:nonstationary}, this implies 
\begin{align*}
f^{(I)}(z;s) = \lim_{\kappa\to 0} \frac{f^{(II)}(z;s,\kappa)}{C^{(II)}(s,\kappa)},\quad 
 \cE^{(I)}(s)=\cE_0(s)-\lim_{\kappa\to 0}\ii \frac{\pi}{\ell^2}\frac{\partial_\tau C^{(II)}(s,\kappa)}{C^{(II)}(s,\kappa)}. 
\end{align*}
Variant II is interesting since the function $f^{(II)}(z;s,\kappa)$ is less complicated than the function $f^{(I)}(z;s)$: by solving the non-stationary equation with the condition $\cE=\cE_0$, one obtains simpler formulas. 

\subsection{Generalizations} The solution algorithm in Sections~\ref{sec:P1}--\eqref{sec:P3} can be generalized to arbitrary variable numbers $N$ (Langmann (2000) \cite{langmann2000}). Moreover, it is possible to obtain explicit formulas for the expansion coefficients (which generalize $a^{(I)}_{n,k}$) and eigenvalues (Langmann (2014)). A technical problem with this algorithm are resonances: multivariable generalization of the energy differences $\cE^{(n)}-\cE_0$ can vanish even for non-zero $n$, and this complicates the algorithm significantly; in particular, due to this resonance problem, it was not possible to prove convergence of these perturbative eigenfunctions for $N\geq 3$ (Langmann (2014)). A way to avoid this resonance problem is to use a linear combination of the commuting operators rather than the Hamiltonian --- this was used  by Langmann, Noumi \& Shiraishi (2022) \cite{langmann2022} to construct a perturbative solution of the elliptic Ruijsenaars model which allowed for a  proof of convergence.  For the non-relativistic elliptic CMR models, there is another perturbative solution due to Komori and Takemura (2002) \cite{komori2002}; this approach is based on standard Kato-Rellich perturbation theory starting with the known solution in the trigonometric case, and it allowed to prove convergence.

Variant II of the solution algorithm providing the solution $f^{(I)}(z;p,\kappa)$ of the non-stationary eCS equation was proposed by Atai \& Langmann (2018) \cite{atai2018}. 
We mention this since Shiraishi (2019) \cite{shiraishi2019} recently presented fully explicit formulas(!) for the $N$-variable generalization of these solutions. 
Shiraishi obtained this formula as a limit of a fully explicit formula for non-stationary Ruijsenaars functions in terms of {\em Nekrasov factors} which, as he conjectured, are solutions of a non-stationary Ruijsenaars equation --- the latter is a $\kappa$ deformation of the eigenvalue equation of the elliptic Ruijsenaars model which, at this point, is not known (it could be related to DELL; see Mironov \& Morozov (2024) \cite{mironov2024}). 
We find these conjectures spectacular: in the least, they give confidence that fully explicit and mathematically beautiful solutions of elliptic CMR models are possible. 
Moreover, this conjecture suggests that, paradoxically, in the context of elliptic CMR models, it can be easier to solve more general models (depending on more parameters). 
If would be interesting to find the non-stationary Ruijsenaars equation having Shiraishi's functions as exact solution. 

\section{Final remarks}\label{sec:remarks} 
We gave an introduction to elliptic CMR systems and the special functions defined by them, aiming at non-experts. 
Different from other such texts in the literature, we emphasized three aspects: (i) the existence of deformed and generalized variants of these models, (ii) kernel functions as  a powerful tool to construct the pertinent special functions, (iii) existence of a scalar product promoting pertinent eigenfunctions to an orthogonal Hilbert space basis. Our motivation to emphasize (i) is that, to make progress in understanding, it is often helpful to try to generalize --- it therefore is important to know in which directions generalizations are possible.  We emphasized (ii) to offer an answer to the questions why we can have confidence that elliptic CMR systems have exact analytic solutions (which in many cases remain to be found): while quantum integrability certainly can give such confidence, there are elliptic CMR systems for which quantum integrability has not yet been proved. However, it seems that kernel functions always exists for such systems, and they provide key ingredients for at least two routes towards solving these systems: solutions by integrals, and perturbative solutions. Until recent, (iii) seemed not possible for the deformed and generalized elliptic CMR systems, but it is a fundamental part for an interpretation as a generalized quantum mechanical model; due to recent results, we can have confidence that such scalar products exist in all cases. 

 This paper is by no means a comprehensive review, and we apologize for many omissions in our reference list.
Instead, we  mention below a few references which, as we hope, can help interested readers to trace the literature on pertinent results which we do not cover.

An earlier and more comprehensive review of elliptic CMR models and the special functions defined by them was given by Ruijsenaars (2004) \cite{ruijsenaars2004}. 
Nekrasov \& Shatiashvili obtained explicit results about the solutions of elliptic CMR models, including the eCS and the Ruijsenaars models, using four dimensional $\mathcal{N}=2$ supersymmetric gauge theory; see Nekrasov \& Shatiashvili (2011) \cite{nekrasov2011} for a review.  Another method used to obtain results about solutions of elliptic CMR models is {\em Separation of Variables}; see Sklyanin (1995) \cite{sklyanin1995} for an introduction. A beautiful subject related to elliptic CMR models is the theory of {\em elliptic hypergeometric functions} with important contributions by van~Diejen, Kajihara, Noumi, Rains, Rosengren, Schlosser, Spiridonov, Warnaar (and others); see the last chapter in the book by Gasper \& Rahman (2011) \cite{gasper2011}  for an introduction and Spiridonov (2007) \cite{spiridonov2007} for a discussion of the relation to elliptic CMR models. We also should mention constructions of eigenfunctions of elliptic CMR models using geometric representation theory, including works by Braverman (2006) and Negut (2011); see Awata {\em et.al} (2023) and references therein. 
To conclude, we mention a recent construction of eigenfunctions of a discrete version of the elliptic Ruijsenaars model by van Diejen \& G\"orbe (2022) \cite{vandiejen2022}, and a recent construction of the generalized eCS model by Kimura \& Lee (2024) \cite{kimura2024} using an approach known as gauge origami.

\bigskip 

\noindent {\bf Acknowledgements:} We would like to thank Masatoshi Noumi and Junich Shiraishi for encouragement and helpful discussions. 
We are grateful to Valdemar Melin, Hjalmar Rosengren, and an anonymous referee for helpful remarks on the manuscript. 
E.L. acknowledges support from the Swedish Research Council, Grant No. 2023-04726. 

\appendix 

\section{Complementary remarks}\label{app:C} 
Remark  (\ref{rem:1})  etc.\ below is referred to as Remark~\ref{app:C}.\ref{rem:1} etc.\ in the main text.   
\begin{enumerate} 
\item\label{rem:1} Since the Jack polynomials and the model first solved by Sutherland were introduced independently nearly at the same time (in Moser's seminal paper, Sutherland's model is one of three examples), this A-type  trigonometric quantum Calogero-Moser system is known as the {\em Calogero-Sutherland model} in the physics literature. It is interesting that history repeated itself: the Macdonald polynomials and the trigonometric Ruijsenaars model were also introduced  independently nearly at the same time. 

\item\label{rem:wp1} One can show that 
\begin{align*}
\sum_{n\in\Z} \frac{(\pi/2\delta)^2}{\sinh^2(\pi(x-2n\ell )/2\delta)} 
\end{align*} 
is equal to $\wp_1(x)$ up to an additive constant. 

\item\label{rem:RandS1} This replacing of the real line $\R$ by a circle parametrized by the interval $[-\ell,\ell]$ is technically convenient for several reasons; for example, the continuous spectrum of a Schr\"odinger operator becomes a pure point spectrum, and eigenfunctions describing scattering solutions become bound states which are much easier to treat.
 
\item\label{rem:IIandIII} There are eigenfunctions of CMR models which do not have a direct physics interpretation in quantum mechanics and which still are interesting from a mathematics points of view; see e.g.\ Halln\"as (2024) \cite{hallnas2024}. 

\item\label{rem:g} One can extend to $g\geq 0$ and still have square-integrable eigenfunctions, but for $0\leq g<\hbar$ the physics interpretation of the model becomes subtle. 
From a mathematical point of view, it is also interesting to consider the case $g<0$ when the eigenfunctions are not square-integrable. 

\item\label{rem:coupling} One writes the coupling constant as $g(g-\hbar)/m$ since the eigenfunctions of $H$ are specified by the coupling parameter $g$ rather than $g(g-\hbar)/m$: 
since $g(g-\hbar)=(\hbar-g)(-g)$, the coupling parameters $g$ and $\hbar-g$ specify two different solutions corresponding to the same coupling constant. 

\item\label{rem:lambda} We note that, due to translational invariance, Jack and Macdonald polynomials have the property 
\begin{align*}
P_\lambda(z)=(z_1z_2\cdots z_N)^{-k}P_{\lambda_1+k,\lambda_2+k,\ldots,\lambda_N+k}(z)\quad (k\in\Z) , 
\end{align*} 
and this is true in the elliptic cases as well. Using this, one can define these functions for all integer vectors $\lambda=(\lambda_1,\ldots,\lambda_N)$ such that 
$$\lambda_1\geq \lambda_2\geq \cdots \geq \lambda_N,$$
allowing also negative $\lambda_N$ as well. This extension is important from  a physics point of view: one has to allow for $\lambda_N<0$ to get a complete set of eigenfunctions; from a mathematics point of view and in the trigonometric cases, it is convenient to restrict to $\lambda_N\geq 0$ since otherwise one gets functions which are not polynomials (but related to polynomials in a simple way). In the elliptic cases, one does not have polynomials anyway, and then there is no longer a good reason for restricting to $\lambda_N\geq 0$. 

\item\label{rem:product} This result is proved in the elliptic Ruijsenaars case (Langmann, Noumi \& Shiraishi \cite{langmann2022}).

\item\label{rem:LameLame1} This is easy to see in the limit $\ell\to\infty$. 

\item\label{rem:eGamma} This elliptic Gamma function was introduced by Ruijsenaars in another form. The form in \eqref{eGamma} is due to Felder \& Varchenkov. 

\item\label{rem:open} There is an interesting proposal for such a generalization by Felder \& Varchenkov called $q$-KZB equation. In Section~\ref{sec:perturbative}, we discuss a result by Shiraishi (2019) suggesting the existence of a non-stationary Ruijsenaars equation which probably is different from the $q$-KZB equation. 

\item\label{rem:Egen} For lack of a better name, we call the parameter $E$ in non-stationary CMR equations \eqref{nSchEq} {\em generalized eigenvalue} --- since $E$ is allowed to depend on $\tau$, it is not an eigenvalue in the conventional sense. 

\item\label{rem:g=-n} As will be seen, these eigenfunctions $\psi(x)$ are proportional to $\vartheta_1(x)^{-n}$, which is why they correspond to $g=-n$ rather than $g=n+1$.

\item\label{rem:constant}  This constant can be computed, but we ignore it for simplicity. 

\item\label{rem:EK} This is Theorem~5.1 in Etingof \& Kirillov (1994) \cite{etingof1994}; note that  the conventions there are $\ell=1$, $2/\kappa_{\mathrm{EK}}=8/\kappa$, and our $\xi$ is equal to $-\ii\pi\lambda/\ell$ there. 

\item\label{rem:saddle} To see this, note that the $\kappa$-dependent part of the integrand can be written as  $\exp( 8G(t)/\kappa)$ with 
$$
G(t) = \sum_{j=1}^n\left(\xi t_j - n\ln\vartheta_1(t_j) \right) + \sum_{1\leq j<k\leq n} \ln\vartheta_1(t_j-t_k) . 
$$
The saddle point equations can therefore be obtained as $\partial_{t_j}G(t)=0$, i.e., 
\begin{align*} 
\xi-n\frac{\vartheta_1'(t_j)}{\vartheta_1(t_j)} + \sum_{\substack{k=1\\k\neq j}}^n \frac{\vartheta_1^\prime(t_j-t_k)}{\vartheta_1(t_j-t_k)}=0\quad (j=1,\ldots,n);  
\end{align*} 
by adding these equations, one obtains \eqref{xieq}, and by inserting this into the saddle point equations one finds the Bethe equations in \eqref{Betheeqs}.

\item\label{rem:neCS}  It is interesting to note that the solution $\vartheta_1(x)^g$ of the non-stationary Lam\'e equation in \eqref{nLame} for $\kappa=2g$ can be obtained as special case $(N,M)=(2,0)$ of \eqref{KNMid}, setting $x_1-x_2=x$. More generally, setting $(N,M)=(N,0)$, one finds that $\psi_0(x)$ in \eqref{psi0} is a solution of the non-stationary eCS equation for $\kappa=Ng$.

\item\label{rem:contour} Use that $dy=\ell d\xi/\pi\ii \xi$. The integration contour is counter-clock wise around a circle $|\xi|=\exp(\pi \epsilon/\ell)$ in the complex $\xi$-plane and, since this contour is closed and the integrand analytic in some neighborhood of the integration contour, the boundary term in \eqref{BT} vanishes. 
The parameter $\epsilon>0$ should be such that $p<|\xi|/|z_j|<1$ on the integration contour $|\xi|=\exp(\pi \epsilon/\ell)$.

\item\label{rem:p} More specifically, one uses the fact that  $\exp(\ii p(x_1+\cdots +x_N))\psi(x)$ is a solution of the non-stationary eCS equation if $\psi(x)$ is a solution. 

\item\label{rem:minimalistic} Allowing for non-zero $a_{0,k}$ for $k>0$ amounts to changing the normalization of the eigenfunction by a $p$-dependent constant. 

\item\label{rem:resonance} We assume $s_1-s_2\neq \Z$ for this to hold true. The case $s_1-s_2\in \Z$ is a case with resonances --- we ignore this case here for simplicity. 

\item\label{rem:K2} Note that, due to translational invariance, the kernel identity
\begin{align*} 
(H_2(x;g)-H_2(y;g))K(x,y)=0
\end{align*} 
for $K(x,y)=K_{2,2}(x,y;g)$ implies the same kernel identity for 
\begin{align*} 
\tilde{K}_{2,2}(x,y;g) = C \ee^{\ii p(x_1+x_2-y_1-y_2)}K_{2,2}(x,y;g) , 
\end{align*} 
for arbitrary $C\in\C\setminus\{0\}$, $p\in\R$. 
By changing variables to $z_j=\exp(\ii\pi x_j/\ell)$, $\xi_j=\exp(\ii\pi y_j/\ell)$ and using \eqref{KNM} for $N=M=2$ we find
\begin{multline*} 
\tilde{K}_{2,2}(x,y;g) = C \Big(\frac{z_1z_2}{\xi_1\xi_2} \Bigr)^{p\ell/\pi}\vartheta_1(x_1-x_2)^g\left(\frac{\ii(\xi_1/\xi_2)^{-1/2}\theta(\xi_1/\xi_2;g)}{\prod_{j,k=1,2}(\ii(z_j/\xi_k)^{-1/2}\theta(z_j/\xi_k;p)  }\right)^g\\
= \frac{C}{\ii^{3g}}\Big(\frac{z_1z_2}{\xi_1\xi_2} \Bigr)^{p\ell/\pi +g} \vartheta_1(x_1-x_2)^g (\xi_1/\xi_2)^{-g/2}\frac{\theta(\xi_1/\xi_2;g)^{g}}{\prod_{j,k=1,2}\theta(z_j/\xi_k;p)^g}.
\end{multline*} 
We drop the factor $(C/\ii^{3g}(z_1z_2/\xi_1\xi_2)^{p\ell/\pi+g}$ (by choosing $C/\ii^{3g}=1$ and $p\ell/\pi+g=0$) to get the result stated in the main text. 

\item\label{rem:R1R2} Since $\theta(z;p)$ is analytic for $p<|z|<1$, the integration contours should be such that $p<|\xi_1/\xi_2|<1$ and $p<|z_i/\xi_j|<1$, i.e., $pR_2<R_1<R_2$ and $1<R_j<1/p$ (we assume $|z_i|=1$); this is implied by $1<R_1<R_2<1/p$. 

\item\label{rem:gauge} To see this, note that one can solve \eqref{Ank} with the conditions \eqref{acond2} even for non-zero $\kappa$, but then one has to allow for non-zero $\cE_{k\geq 0}$: one can regard \eqref{acond2} and \eqref{Econd} as two different gauge conditions one can use to solve \eqref{Ank} for non-zero $\kappa$. Using \eqref{acond2}, the recursive solution of \eqref{Ank} is exactly like in Section~\ref{sec:VariantI} (Variant~I), and one obtains a solution $f^{(I)}(z;s,\kappa)$, $\cE^{(I)}(s,\kappa)$; in the limit $\kappa\to 0$, this reduces to the solutions $f^{(I)}(z;s)$, $\cE^{(I)}(s)$ in Section~\ref{sec:VariantI}. At non-zero $\kappa$, the solutions $f^{(I)}(z;s,\kappa)$ and $f^{(II)}(z;s,\kappa)$ are equal up to normalization (depending on $\tau$); the relative constant is determined by the constant parts $C^{(I)}(s,\kappa)=1$ and $C^{(II)}(s,\kappa)$: $f^{(I)}(z;s,\kappa)=f^{(II)}(z;s,\kappa)/C^{(II)}(s,\kappa)$. 

\end{enumerate}  

\section{Elliptic functions}\label{app:ell} 
We spell out how the functions $\wp_1(x)$ in \eqref{wp1} and $\vartheta_1(x)$ in \eqref{theta1} are related to the ones defined by Whittaker \& Watson (1920) \cite{whittaker1920} --- the latter reference is referred to as WW in the rest of this appendix. We also collect some well-known identities used in the main text. 
 
For the $\wp_1$-function:  
\begin{align*}
\wp_1(x)=\wp(x)+\frac{\eta_1}{\omega_1} 
\end{align*} 
where 
\begin{align*}
\frac{\eta_1}{\om_1}=\frac{\zeta(\omega_1)}{\om_1} = \Big(\frac{\pi}{\om_1}\Big)^2\Big( \frac1{12}-\sum_{n=1}^\infty \frac{p^n}{(1-p^n)^2}\Big),
\end{align*} 
with $\wp(x)=\wp(x|\om_1,\om_2)$ and $\zeta(x)=\zeta(x|\om_1,\om_2)$ the Weierstrass functions as defined by WW with $(\omega_1,\omega_2)=(\ell,\ii\delta)$. 

For the theta function: 
\begin{align*} 
\vartheta_1(x)=\vartheta^{\mbox{\tiny{WW}}}_1(\pi x/\ell|\tau)/C(\tau)
\end{align*} 
with $\vartheta^{\mbox{\tiny{WW}}}_1(z|\tau)$ 
the theta function as defined by WW,  where $\tau=\om_2/\om_1=\ii\delta/\ell$ and 
\begin{align*} 
C(\tau)=\ee^{\ii\pi\tau/4}\prod_{n=1}^\infty(1-\ee^{2n\ii\pi\tau}). 
\end{align*} 

While $\wp_1(x)$ is doubly periodic: $\wp_1(x+2\ell)=\wp_1(x+2\ii\delta)=\wp_1(x)$, the theta function $\vartheta_1(x)$ has the following quasi-periodicity properties:
\begin{align}\label{vtheta1qper}
\vartheta_1(x+2\ell) = \vartheta_1(x),\ \ \ \vartheta_1(x+2\ii\delta) = -\ee^{\pi\delta/\ell}\ee^{-\ii\pi x/\ell}\vartheta_1(x).
\end{align}

The theta function $\vartheta^{\mbox{\tiny{WW}}}=\vartheta^{\mbox{\tiny{WW}}}_1(z|\tau)$ in WW satisfies the heat equation: 
\begin{align*}
\tfrac{1}{4}\pi\ii\frac{\partial^2}{\partial z^2}\vartheta^{\mbox{\tiny{WW}}}_1 + \frac{\partial}{\partial\tau}\vartheta^{\mbox{\tiny{WW}}}_1=0. 
\end{align*} 
Due to our different normalization and scaling of $\vartheta_1(x)$, the heat equation obeyed by our theta function is slightly more complicated; we write it as 
\begin{align}\label{heat} 
\Big( \ii \frac{\pi}{\ell^2}\frac{\partial}{\partial\tau} -\frac{\partial^2}{\partial x^2} -c_0 \Bigr)\vartheta_1(x)=0
\end{align} 
with the constant (note that $\om_1=\ell$)
\begin{align} 
c_0 = - \ii \frac{\pi}{\ell^2}\frac{\partial_\tau C}{C}=\Big(\frac{\pi}{\ell}\Big)^2\Big(\frac14 -2\sum_{n=1}^\infty\frac{np^n}{1-p^n}\Big) = 2\frac{\eta_1}{\omega_1} + \Big(\frac{\pi}{\ell}\Big)^2\frac1{12}. 
\end{align}

We also need the following series representation of the Weierstrass $\wp$-function
\begin{align}\label{wpseries} 
\wp_1(x) = \frac{(\pi/2\om_1)^2}{\sin(\pi x/2\om_1)^2} -2(\pi/\om_1)^2\sum_{m=1}^\infty \frac{mp^m}{1-p^m}\cos(m\pi x/\om_1); 
\end{align} 
see WW. Changing variables to $z=\exp(\ii \pi x/\ell)$ we get $2\cos(m\pi x/\om_1) = z^m+z^{-m}$ and 
$$\sin^2(\pi x/2\om_1)=-(1/4)(z^{1/2}-z^{-1/2})^2=-(1/4z)(1-z)^2,$$ 
which gives 
$$
\wp_1(x) = -\Big(\frac{\pi}{\om_1}\Big)^2\Big( \frac{z}{(1-z)^2} +\sum_{m=1}^\infty \frac{mp^m}{1-p^m}(z^m+z^{-m}) \Big) .
$$
Expanding two geometric series we obtain \eqref{wpseries}. 

To conclude, we note that 
\begin{align*} 
\zeta_1(x)=\frac{\partial}{\partial x}\ln \vartheta_1(x) = \zeta(x;\om_1,\om_2) -\frac{\eta_1}{\om_1}x
\end{align*}  
is equal to the function $\vartheta_1'(x)/\vartheta_1(x)$ appearing repeatedly in Section~\ref{sec:Bethe}; we avoid using the function $\zeta_1(x)$ in the main text for pedagogical reasons.
\section{Classification of elliptic CMR models}\label{app:CMR} 
We give some remarks on the classification of elliptic CMR models, together with an overview of the models we discuss in the main text. 

CMR systems can be classified by (i) whether the interactions are given by rational, hyperbolic, trigonometric or elliptic functions, (ii) by a root system like $A_{N-1}$ or $BC_N$ (this goes back to Olshanetsky \& Perelomov),\footnote{The CMR models related to the root systems $B_N$ and $C_N$ are special cases of the $BC_N$ models. There are also CMR systems related to $D_N$ and the exceptional root systems (like $E_8$ or $G_2$) which we do not discuss; see  Komori (2001) \cite{komori2001} for such models in the relativistic case.} and (iii) whether the model is non-relativistic (i.e., Calogero-Moser type) or relativistic (i.e., Ruijsenaars-van Diejen type); see e.g.\ Halln\"as (2024) \cite{hallnas2024} (we discuss (i) and (iii) in Section~\ref{sec:intro}). Different CMR models can be named systematically using this classification (i)--(iii), but we use other names collected in the list below (in the literature,  the name {\em quantum Calogero-Moser} is often used instead of our {\em non-relativistic CMR}, and {\em relativistic quantum Calogero-Moser} is used instead of our {\em relativistic CMR}): 

\begin{itemize} 
\item eCS model = $A_{N-1}$ elliptic non-rel.\ CMR model 
\item elliptic Ruijsenaars model = $A_{N-1}$ elliptic rel.\ CMR model 
\item quantum Inotzemtsev model = $BC_N$ elliptic non-rel.\ CMR model 
\item elliptic van-Diejen model = $BC_N$ elliptic rel.\ CMR model 
\item Lam\'e equation = $A_{1}$ elliptic non-rel.\ CMR model 
\item Heun equation = $BC_{1}$ elliptic non-rel.\ CMR model. 
\end{itemize} 

CMR models allow for a generalization in the elliptic case (and not in the other three cases) which we distinguish from the standard case by adding the word {\em non-stationary}, e.g.\ non-stationary eCS equation etc.; the motivation for this name is that this generalizations looks similar to a non-stationary Schr\"odinger equation (even though the time variable $\tau$ is imaginary --- physicists might prefer to think of it as generalized heat equations); these generalizations are often called KZB equation in the literature. 

All CMR models have generalizations distinguished from the standard model by adding the word {\em deformed} to the name, e.g., deformed eCS model etc. (see e.g.\ Sergeev \& Veselov (2004) \cite{sergeev2004}). 
These deformed CMR models describe interacting systems with arbitrary numbers of two types of particles. The deformed CMR models are related to root systems of Lie superalgebras and, for this reason, the eigenfunctions arising in these models are distinguished from the standard ones by adding the word {\em super}; there exist, for example, {\em super Jack polynomials} and {\em super Macdonald polynomials} which are well-understood through works by Sergeev \& Veselov and others. In the elliptic case, the deformed CMR models allow for a further generalization describing arbitrary numbers of four particles types and which we propose to distinguish from the standard models by adding the word {\em generalized} to the name, e.g. {\em generalized eCS model} etc.

\end{document}